\documentclass[aip,amsmath,amssymb,reprint]{revtex4-2}     

\usepackage{graphicx}           
\usepackage{amssymb}
\usepackage{amsmath}
\usepackage{natbib}
\usepackage{color}
\usepackage{eucal}
\usepackage{url}
\usepackage{hyperref}

\newcommand{\kon}{k_\mathrm{on}}
\newcommand{\koff}{k_\mathrm{off}}

\begin{document}

\title{Sliding across a surface: particles with fixed and mobile ligands }
\author{Janna Lowensohn}
\thanks{These two authors contributed equally}
\affiliation{ 
Center for Nonlinear Phenomena and Complex Systems, Code Postal 231, Universit\'e Libre de Bruxelles, Boulevard du Triomphe, 1050 Brussels, Belgium
}%
\author{Laurie Stevens}
\thanks{These two authors contributed equally}
\affiliation{ 
Interuniversity Institute of Bioinformatics in Brussels, ULB-VUB, La Plaine Campus, 1050 Brussels, Belgium
}%
\affiliation{ 
Center for Nonlinear Phenomena and Complex Systems, Code Postal 231, Universit\'e Libre de Bruxelles, Boulevard du Triomphe, 1050 Brussels, Belgium
}%
\affiliation{ 
Applied Physics Research Group, Vrije Universiteit Brussel, Pleinlaan 2, 1050 Brussels, Belgium 
}
\author{Daniel Goldstein}
\affiliation{ Department of Physics and Astronomy, Tufts University,
574 Boston Avenue, Medford, Massachusetts 02155, USA
}
\author{Bortolo Matteo Mognetti}
\email{Bortolo.Matteo.Mognetti@ulb.be}
\affiliation{ 
Center for Nonlinear Phenomena and Complex Systems, Code Postal 231, Universit\'e Libre de Bruxelles, Boulevard du Triomphe, 1050 Brussels, Belgium
}%


\date{\today}

\begin{abstract}

A quantitative model of the mobility of ligand presenting particles at the interface is pivotal to understanding important systems in biology and nanotechnology. In this work, we investigate the emerging dynamics of particles featuring ligands that selectively bind receptors decorating an interface. The formation of a ligand-receptor complex leads to a molecular bridge anchoring the particle to the surface. We consider systems with reversible bridges in which ligand-receptor pairs bind/unbind with finite reaction rates. For a given set of bridges, the particle can explore a tiny fraction of the surface as the extensivity of the bridges is finite. 
We show how, at time scales longer than the bridges’ lifetime, the average position of the particle diffuses away from its initial value. We distill our findings into two analytic equations for the sliding diffusion constant of particles carrying mobile and fixed ligands. We quantitatively validate our theoretical predictions using reaction-diffusion simulations. We compare our findings with results from recent literature and discuss the molecular parameters that likely affect the particle's mobility most. Our results, along with recent literature, will allow inferring the microscopic parameters at play in complex biological systems from experimental trajectories.

\end{abstract}

\maketitle

\section{Introduction}
Ligand--receptor interactions underly most of the biological processes found at the cell membrane, like adhesion and signaling \cite{Alberts}.  
Much work has quantified the interaction strength between particles (e.g., colloids, viruses, or vesicles) and surfaces (e.g., supported lipid bilayers or cell membranes) mediated by ligand--receptor complexes \cite{mognetti2019programmable,KitovJACS2003,licata2008kinetic,Martinez-Veracoechea_PNAS_2011,curk2016design,A-UbertiNPJ2017}. 
Different groups have also studied the particle--surface first contact leading to the formation of an interacting patch  \cite{shenoy2005growth,zhang2008nucleation,atilgan2009nucleation}.
Despite the particle's mobility being pivotal in many biological processes, except for studies in a shear flow \cite{HAMMER1987475,dasanna2017rolling,porter2021shear}, little has been done to understand the rolling/sliding dynamics of a functionalized particle at the interface \cite{olah2013superdiffusive,C8SM01430B,jana2019translational,marbach2021nanocaterpillar,korosec2021substrate}. 
For instance, successful cellular invasions by Influenza A Viruses \cite{de2020influenza,vahey2019influenza} and Herpes Viruses \cite{Delgusteeaat1273} require the invader to diffuse along the cell membrane while remaining bound.
Similarly, it is believed that the Malaria parasite uses ligand gradients to reorient itself towards a configuration favoring erythrocyte invasion \cite{dasgupta2014membrane}. In the self--assembly of functionalized colloids, the mobility of bound particles is pivotal to relax disordered aggregates into crystalline structures \cite{wang2015crystallization}.

\begin{figure}[h]
\includegraphics[width=250pt]{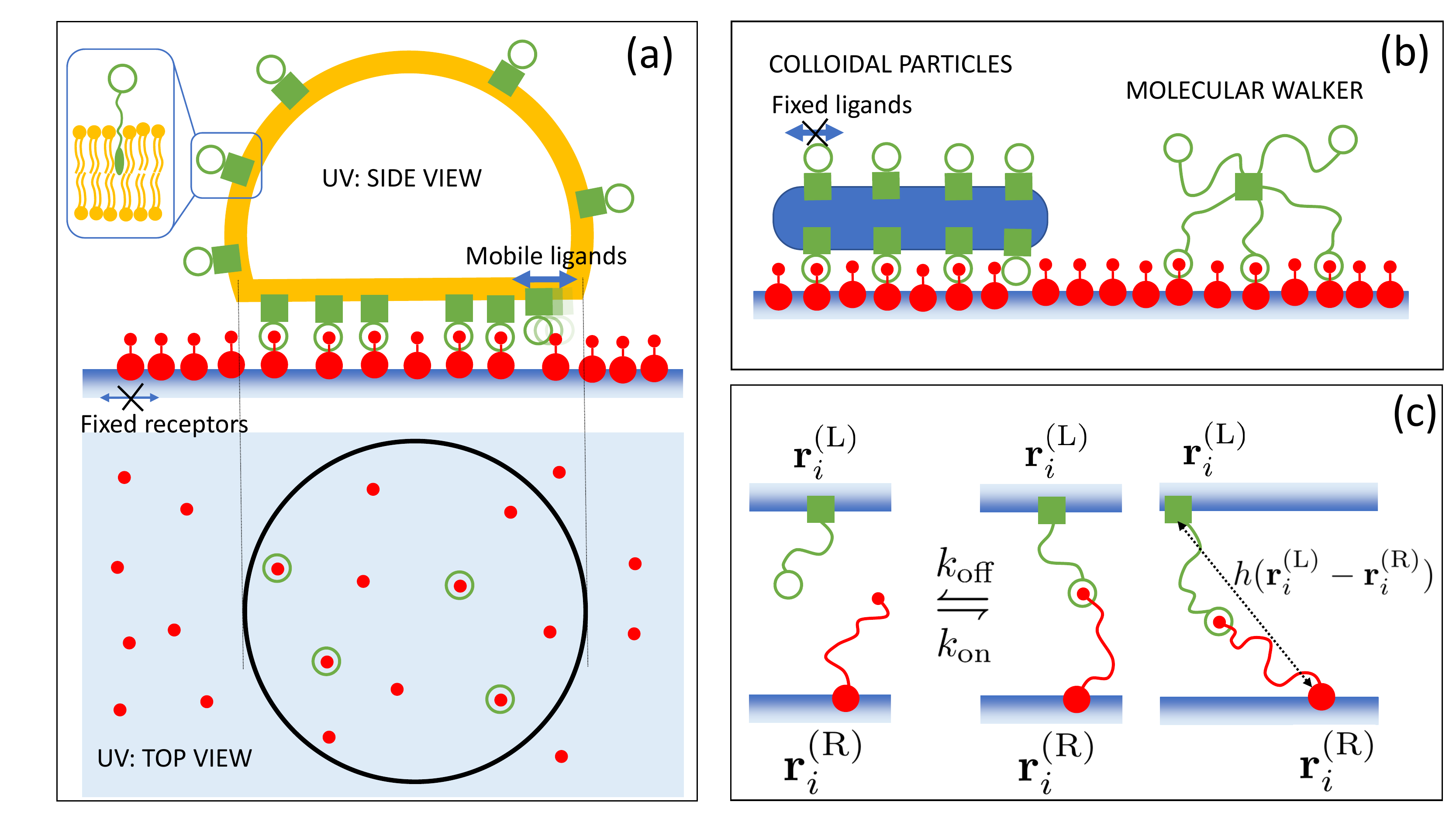}
\caption{{\em (a), (top)} Unilamellar Vesicles (UVs) carry ligands anchored to the lipid bilayer. For liquid bilayers, ligands are mobile along the UV surface. {\em (bottom)} The contact region between the vesicle and the surface is represented by a 2D disk shown in black. Bridges form and break inside the contact region. {\em (b), (left)} Colloidal particles carrying ligands with fixed tethering points. {\em (right)} In molecular walkers, ligands are the legs of star-shaped polymers tipped by reactive complexes (e.g., Ref.~\onlinecite{pei2006behavior}). {\em (c), (left)} Bridges form and break with a rate constant equal to, respectively, $\kon$ and $\koff$. {\em (right)} The bond potential $h$ controls the stretchability of the bridges. {\em (a, b, c)} The different elements of the figure are not to scale.}\label{Fig:LR} 
\end{figure}

This study aims at improving our understanding of how particles bound to receptor--expressing surfaces \footnote{Without loss of generality, in this contribution ligands and receptors refer to linker molecules tethered, respectively, to the particles and the surface.} can laterally diffuse (slide).
We develop a model in which the particle's center of mass is constrained to remain within a finite area, $\Omega$, by the presence of bridges. 
$\Omega$ is entirely determined by a subset of bridges (in the following constraining bridges). The emerging dynamics are then controlled by the timescale over which a constraining bridge is added or removed along with the average displacement of the particle following an update of $\Omega$. We summarize our results using two simple equations (Eqs.~\ref{Eq:D}, \ref{Eq:Dfixed}) expressing the diffusion constant in terms of the single-bridge off--rate $\koff$,  the average number of bridges $\langle n_b\rangle$, and the area of the surface reachable by a single ligand.
Our predictions hold in the limit in which the evolution of $\Omega$ is slow as compared to the timescales taken by the particle to explore $\Omega$. We clarify how this approximation holds based on the finding that the number of constraining bridges remains constant in the many-bridge limit ($\langle n_\mathrm{cb}\rangle \approx 5$). In a recent contribution, Kowalewski et {\em al.}  derived an expression similar to Eq.~\ref{Eq:D} (Eq.~\ref{Eq:DK}) for the diffusion of a molecular walker (Fig.~\ref{Fig:LR}b) \cite{Parker2005}. As compared to Ref.~\onlinecite{Parker2005}, the present contribution clarifies how Eqs.~\ref{Eq:D} and \ref{Eq:DK} hold only if  $\langle n_\mathrm{cb}\rangle$ does not increase with the number of bridges.

In Sec.~\ref{Sec:Model} we discuss our model as compared to experimental systems. In Sec.~\ref{Sec:Theory}, we derive the emerging diffusion constant for particles with mobile ligands (Eq.~\ref{Eq:D}). In Sec.~\ref{Sec:Simulations}, we validate the results of Sec.~\ref{Sec:Theory} using reaction--diffusion simulations \cite{jana2019translational}. In Sec.~\ref{Sec:fixed}, we present our predictions for fixed ligands (Eq.~\ref{Eq:Dfixed}). Finally, in Sec.~\ref{Sec:Discussion}, we summarize and discuss the findings of the paper in view of recent results.\cite{marbach2021nanocaterpillar,Parker2005}

\section{The model system}\label{Sec:Model}
Unilamellar vesicles are a class of particles usually employed to carry mobile ligands (Fig.\ \ref{Fig:LR}). Ligands (like membrane proteins or synthetic moieties) are anchored to the membrane through trans-membrane domains, hydrophobic complexes (inset of Fig.~\ref{Fig:LR}a, top), or covalent (e.g., glycosidic) bonds \cite{mognetti2019programmable}.
A schematic model of functionalized vesicles at the interface would comprise a contact region and an outer cap (Fig.~\ref{Fig:LR}a). Ligand-receptor bridges can form only inside the contact region, in the following modeled by a disk (Fig.~\ref{Fig:LR}a, bottom). The outer region acts as a finite reservoir of ligands. Neglecting fluctuations in the direction orthogonal to the surface, we model the system using a 2D representation in which the particle is identified with the disk corresponding to the contact region (Sec.~\ref{Sec:static_bridges}). This model does not track the position of the mobile, free ligands which are mapped into uniform, depletable densities (as the total number of ligands is fixed).

Particles with fixed ligands (Fig.~\ref{Fig:LR}b, left) include colloids functionalized by synthetic moieties or biological particles like viruses (e.g., Ref.~\onlinecite{vahey2019influenza}). Spherical particles tend to roll rather than slide \cite{C8SM01430B,jana2019translational}. Sliding is occasionally more prominent in non-spherical particles like, e.g., the rod--shaped strand of the Influenza A virus. In Sec.~\ref{Sec:fixed}, we predict the sliding diffusion constant using a 2D model (Fig.~\ref{Fig:Fixed_Model}).

Ref.~\onlinecite{Parker2005} employed a 2D model similar to the one of Fig.~\ref{Fig:LR}a, bottom to study molecular walkers (Fig.~\ref{Fig:LR}b, left). Molecular walkers are made from star-shaped polymers tipped by reactive moieties (e.g., Ref.~\onlinecite{pei2006behavior}). Each ligand can then reach out to receptors found inside the circle centered over the projection of the branching point onto the surface with a radius equal to the length of the ligand.

Fig.~\ref{Fig:LR}c reports the important molecular parameters of the system. Bridges form or break with a rate constant equal to $\kon$ or $\koff$, respectively. We refer to Secs.~\ref{Sec:Simulations} and \ref{Sec:fixed} for the calculation of the rates, respectively, for fixed and mobile ligands.
The reaction rates have a major impact on the emerging dynamics. A system composed of static bridges would result in a particle arrested on a surface. Another important element is the bond potential $h$. Ligand/receptor backbones are often constituted by flexible polymers. For ideal polymers, stretching the tethering points far away is contrasted by a harmonic attraction, $h(d)=-k d^2/2$ (where $d$ is the lateral distance between tethering points). For non-ideal polymers, the bond potential is a multi--body function that also depends on neighboring ligands/receptors. For more rigid backbones, the bond potential is dominated by the finite stretchability of the bridges. In this work, for fixed ligands, we use a square--well bond potential,  $h( d)=0$ if $d<\lambda$, $h(d)=\infty$ if $d>\lambda$, where $\lambda$ is the maximal lateral extensibility of a bridge (e.g., \cite{vahey2019influenza}). This choice is mainly motivated by the possibility of deriving universal analytic results which could then be used (when compared with other systems) to assess the importance of the bond potential $h$ (Sec.~\ref{Sec:Discussion}).

For mobile ligands (Fig.~\ref{Fig:LR}a), the bond potential $h$ plays a minor role as averaging over the ligand position results in a null force exerted onto the particle. This consideration holds only if the tethering point anchoring the bridge to the vesicle can explore most of its available configurational space before unbinding.

In this study, we only consider fixed, randomly distributed receptors. We comment on the mobility of the receptors and their distribution in Secs.~\ref{Sec:fixed} and \ref{Sec:Discussion}.

\section{Theoretical predictions of the sliding diffusion constant }\label{Sec:Theory}

To model the emerging dynamics of particles carrying mobile ligands, in the next section (Sec.~\ref{Sec:static_bridges}), we characterize the configurational space ($\Omega$) available to a particle featuring a static set of bridges. In Secs.~\ref{Sec:tau} and \ref{Sec:delta}, we then investigate, respectively, the rate at which $\Omega$ evolves when the set of bridges changes because a bridge is either added or removed and the average displacement of the particle following a change of $\Omega$. Based on such an understanding, in Sec.~\ref{Sec:D} we propose an analytic expression for the emergent diffusion constant $D$ (Eq.~\ref{Eq:D}). Eq.~\ref{Eq:D} holds in the limit in which $D$ is not limited by the diffusion constant of the free particle ($n_b=0$), $D_0$. The simulation protocol employed in Secs.~\ref{Sec:Simulations}, \ref{Sec:fixed} remains reliable at low values of $D_0$.

\begin{figure*}
\includegraphics[width=435pt]{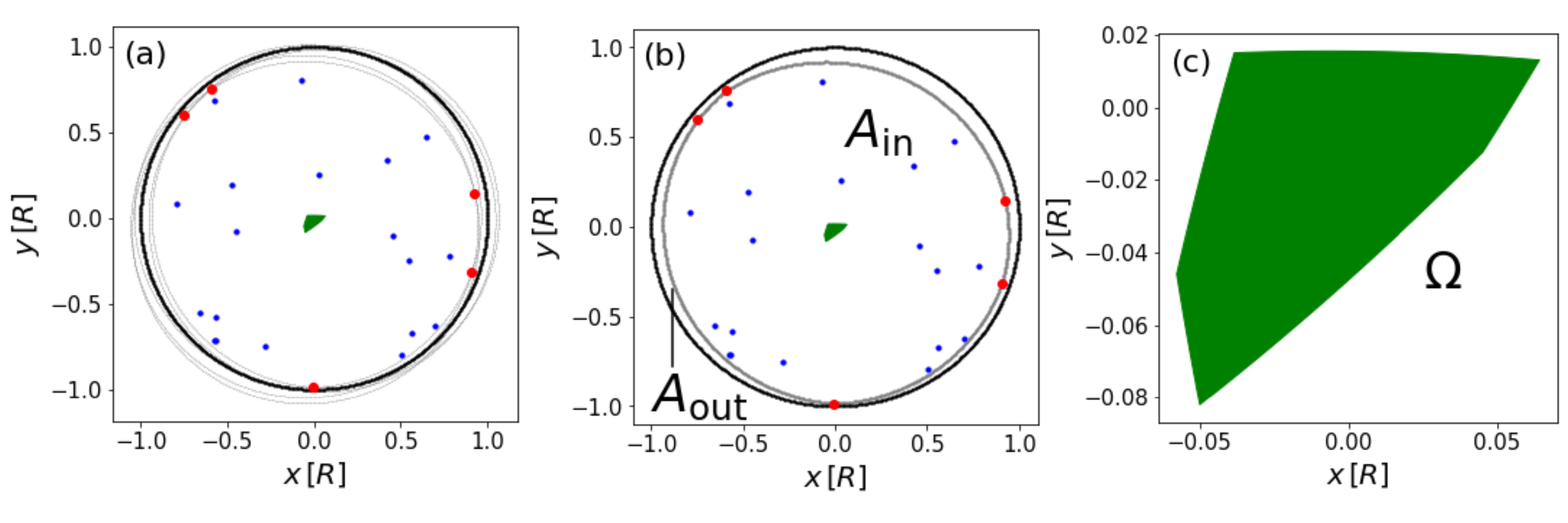}
\caption{({\em a}) CBs (thick, red points) in a system with $n_b=25$ bridges uniformly distributed inside the particle's perimeter (black line). The gray circles represent configurations which simultaneously touch two CBs (such configurations correspond to the vertices of $\Omega$ in Fig.~\ref{Fig:Model}c). ({\em b}) The gray line is the curved polygon with vertices given by the set of CBs. Adding a new bridge inside the curved polygon does not further constrain the particle's position. ({\em c})  The colored region highlights the configurational area ($\Omega$) available to the center of mass for the set of CBs of panel {\em a} and {\em b}. For comparison, $\Omega$ has been inserted in panel $a$ and $b$ (colored regions).  }\label{Fig:Model}
\end{figure*}

\subsection{Constraining bridges (CBs)}\label{Sec:static_bridges}

Following on from Sec.~\ref{Sec:Model} (Fig.~\ref{Fig:LR}a), we model particles functionalized by mobile ligands using 2D disks of radius R moving over a surface decorated by randomly distributed, fixed receptors (Fig.~\ref{Fig:Model}a). The model tracks the position of the disk and the ensemble of receptors (dots in Fig.~\ref{Fig:Model}a) forming a bridge with a ligand moving on the particle.

We assume that bridges cannot be formed with receptors outside the particle's perimeter. 
Therefore, for mobile ligands, we assume that the interaction range is much smaller than the disk's radius ($\lambda\ll R$).
For a given particle position, the bridges are taken as uniformly distributed inside the circle (Fig.~\ref{Fig:Model}a). 
For a given set of bridges, the circle's center can then explore a finite fraction of the surface compatible with the fact that bridges remain inside the perimeter of the particle rattling around the set of bridges (gray circles in Fig.~\ref{Fig:Model}a).

Fig.~\ref{Fig:Model}c highlights the configurational space available to the disk's center ($\Omega$) corresponding to the bridges of Fig.\ \ref{Fig:Model}a. $\Omega$ comprises an ensemble of vertices joined with curved edges (in the following we will refer to such an object as a 'curved polygon'). The edges are arcs of circles of radius $R$ centered on the bridges in contact with the perimeter (thick, red points in Fig.~\ref{Fig:Model}a). Similarly, the vertices of $\Omega$ correspond to configurations in which the perimeter of the particle (gray lines in Fig.~\ref{Fig:Model}a) touches two bridges.
\\
\begin{figure}[h!]
\includegraphics[width=230pt]{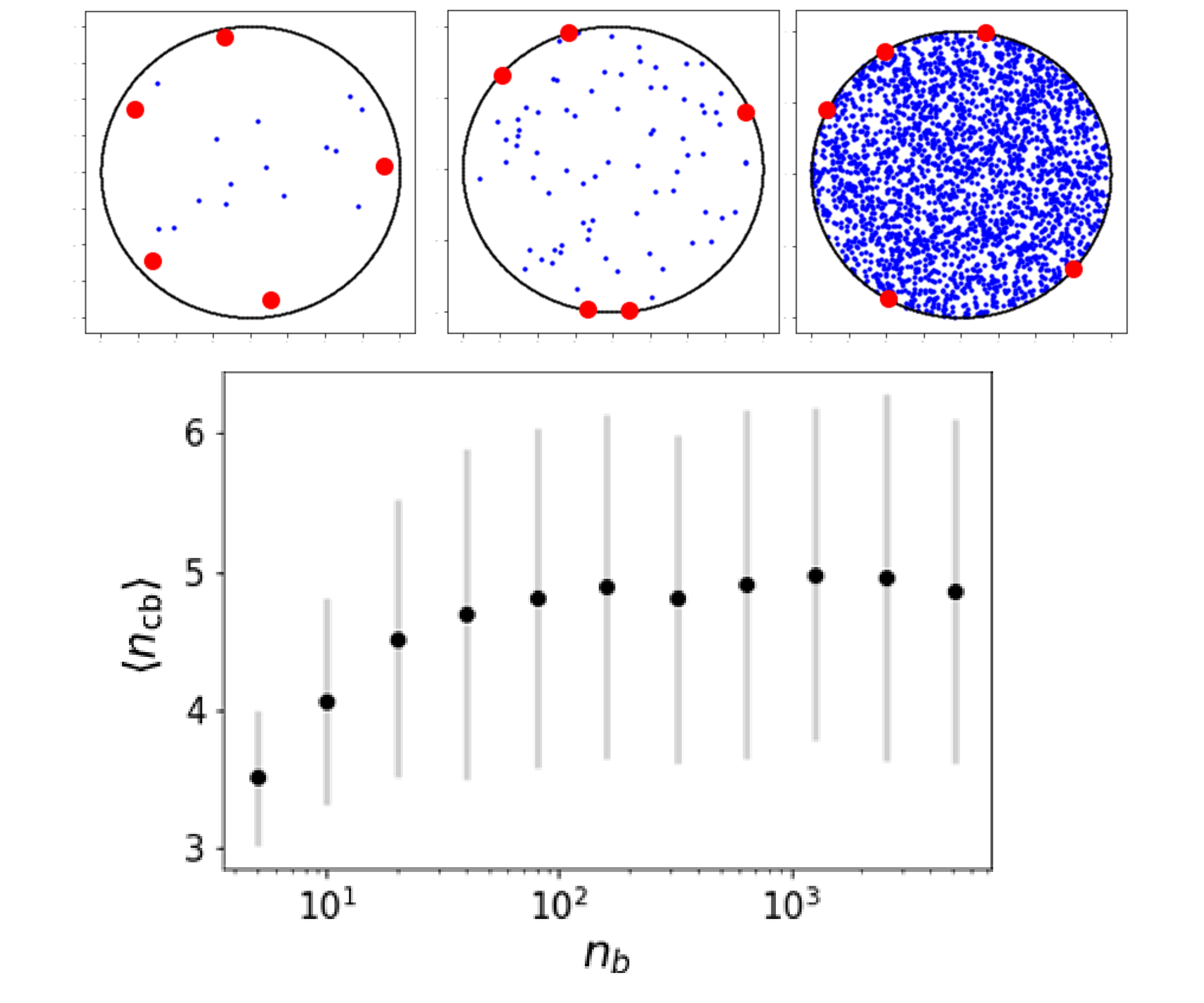}
\caption{(bottom) Average number of constraining bridges (CBs) as a function of the number of bridges. Gray bars represent the variance calculated using $10^3$ independent samples with bridges unifomly distributed inside the disk. The number of bridges does not fluctuate $n_b=\langle n_b \rangle$. (top) Three samples with different number of bridges (from left, $n_b=20$, $n_b=80$, and $n_b=2560$) featuring the same number of CBs (in red). }\label{Fig:ncb}
\end{figure}
%
%
%
%
%
%
In general, only a finite fraction of bridges can come  in contact with the perimeter of the particle (Fig.\ \ref{Fig:Model}a). In the following, we will refer to them as 'Constraining Bridges' (CBs). Intriguingly, the number of CBs ($n_\mathrm{cb}$) remains finite when increasing the number of bridges $n_b$ (Fig.~\ref{Fig:ncb}).
Fig.~\ref{Fig:ncb} reports the average ($\langle n_\mathrm{cb} \rangle$) and the variance of the number of CBs as a function of $n_b$. 
The CBs can be defined as the smallest subset of bridges that, when joined by circle arcs of radius $R$, contain all the remaining $n_b-n_\mathrm{cb}$ points (Fig.~\ref{Fig:Model}b). The sets of CBs are then similar to the vertices of a convex hull \cite{renyi1963konvexe,efron1965convex} with the difference for convex hulls being that the edges between the points on the border are straight lines (SI Fig.~1) rather than the arcs shown in Fig.~\ref{Fig:Model}b.
This key observation justifies the fact that, contrary to what is observed in Fig.~\ref{Fig:ncb}, the number of points belonging to the convex hull diverges logarithmically with $n_b$ \cite{renyi1963konvexe,efron1965convex}.
Notice, for instance, that the stretching of two bridges far away from each other towards the particle's border pins the particle in a single configuration and would impede any other bridge from touching the perimeter (right panel of SI Fig.~1).
\\
\begin{figure}[h!]
\includegraphics[width=230pt]{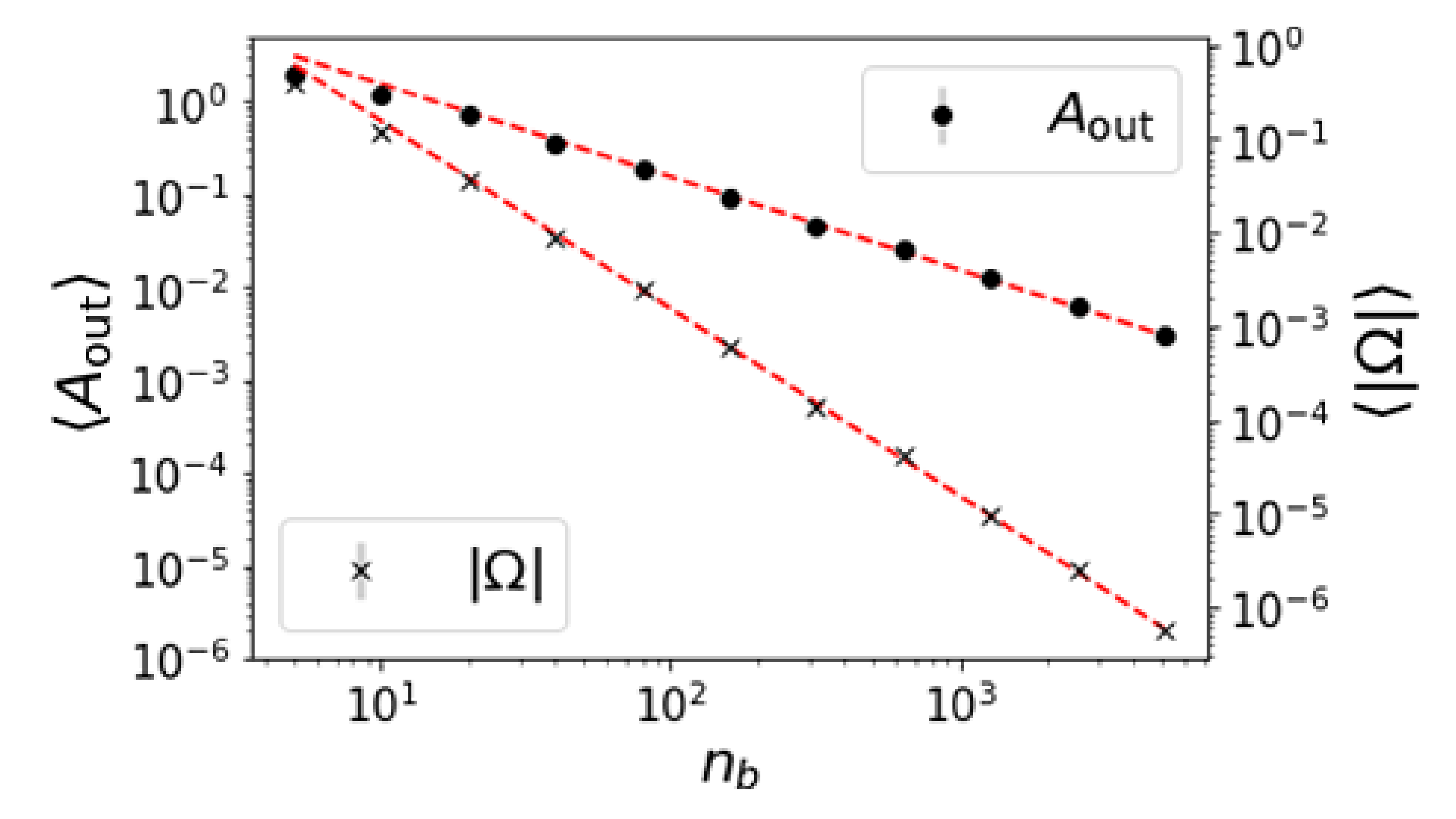}
\caption{Average area of the surface outside the curved polygon ($\langle A_\mathrm{out}\rangle$, Fig.~\ref{Fig:Model}b) and area of the configurational space available to the center of mass of the particle $\langle |\Omega| \rangle$ (Fig.~\ref{Fig:Model}c). Dashed lines are the fitting functions reported in Eqs.~\ref{Eq:AoutTau} and \ref{Eq:Omega}. The errorbars (calculated using 10$^3$ samples as in Fig.~\ref{Fig:ncb}) are smaller than the symbol size. The number of bridges does not fluctuate $n_b=\langle n_b \rangle$. The $y$ scales are in units of $R^2$.}\label{Fig:AoutOmega}
\end{figure}
We define by $A_\mathrm{out}$ the area of the region outside the curved polygon identified by the CBs (Fig.~\ref{Fig:Model}b).
$A_\mathrm{out}$ will be used in the next section to estimate the probability that $\Omega$ changes following the formation of a new bridge.

We now consider the case in which bridges break and form  (corresponding to points that disappear and appear in Fig.~\ref{Fig:Model}a) with a rate constant equal to $k_\mathrm{on}^T$ and $k_\mathrm{off}$, respectively.
We decompose the motion of the particle into two components. At short timescales, the particle rattles around the current CBs. At larger timescales, comparable or larger than the lifetime of a given set of CBs ($\tau_\mathrm{cb}$), the particle is driven by the diffusive motion of the CBs \footnote{Notice that the set of CBs is expected to diffuse as the position of the added and deleted bridges is not the same.}.
Hence, the emerging mobility of the particle is controlled by $\tau_\mathrm{cb}$ and the  typical displacement of the particle following an update of the CBs, $\delta_\mathrm{cb}$. These two quantities are discussed in the next two sections (Secs.~\ref{Sec:tau} and \ref{Sec:delta}).  First, notice that these two dynamics are not strictly independent as possible updates of the CBs depend on the actual position of the particle. Second, this decomposition is valid only in the limit in which the particle can explore an area comparable with $\Omega$ (Fig.~\ref{Fig:Model}c and Fig.~\ref{Fig:AoutOmega}) before the next update of the CBs takes place. 
This is often the case given that the system explores the polygon with a diffusion constant $D_0$ with $D_0=10^6-10^4\,$nm$^2/$s for micrometer-nanometer particles.

\subsection{CBs reconfiguration time ($\tau_\mathrm{cb}$)}\label{Sec:tau}

\begin{figure}[h!]
\includegraphics[width=230pt]{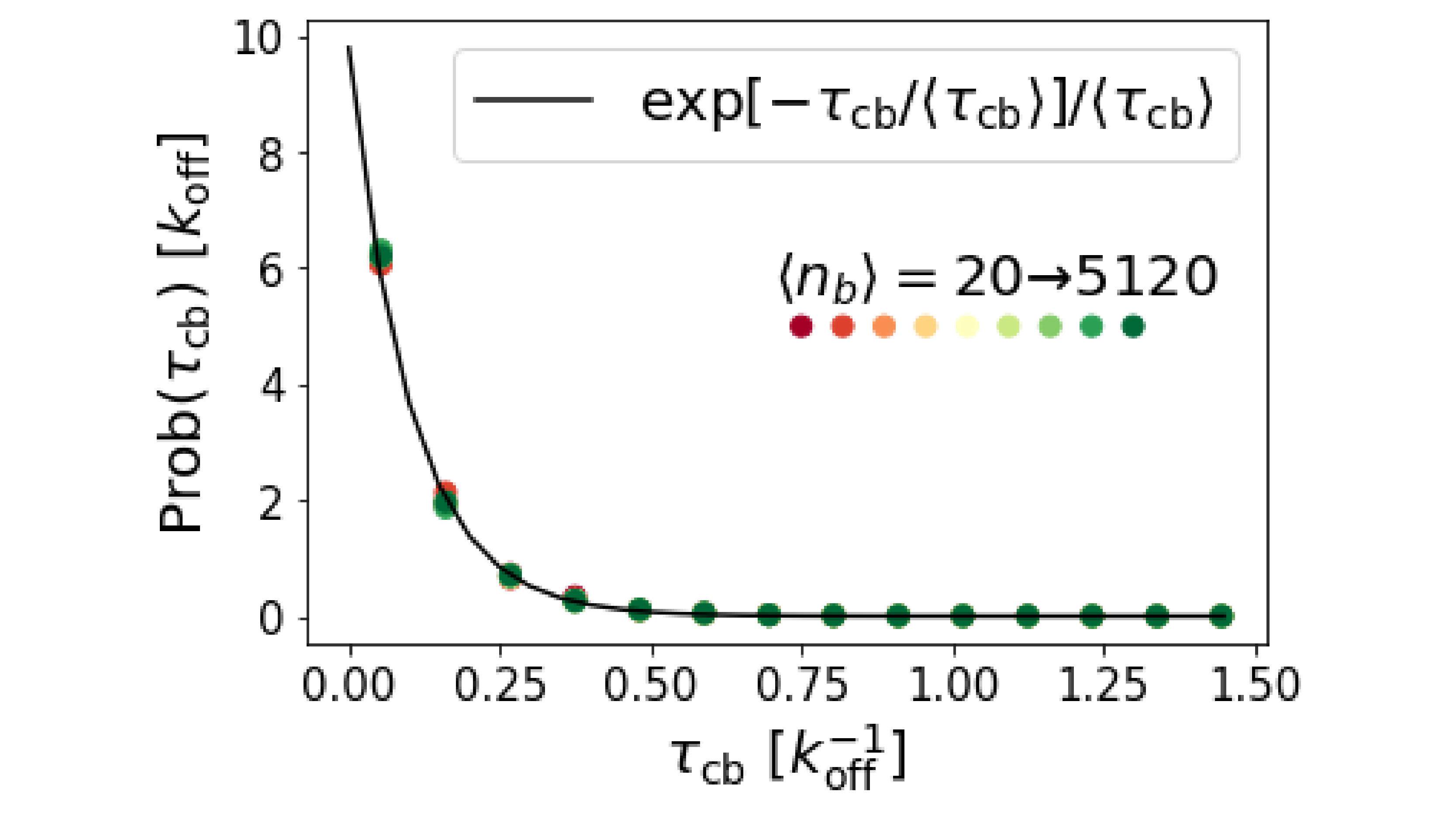}
\caption{Ditribution of the time to update the set of CBs, $\tau_\mathrm{cb}$. Symbols are simulations featuring different average numbers of bridges ($\langle n_b\rangle=20$--$2560$). For each value of $\langle n_b\rangle$, we generate $5\cdot 10^4$ different sets of CBs (SI Sec.~II). The line is a Poisson distribution with average given by Eq.~\ref{Eq:AoutTau} with $\langle n_\mathrm{cb} \rangle=4.9$ (Fig.~\ref{Fig:ncb}). }\label{Fig:P_t_cb}
\end{figure}
There are two possible events leading to a change of the set of CBs (Fig.~\ref{Fig:Model}a) and thus affect $\tau_\mathrm{cb}$. In the first one (event $a$), a free ligand binds a bridge in the region outside the curved polygon ($A_\mathrm{out}$ in Fig.~\ref{Fig:Model}b). In event $b$, a receptor belonging to the CBs unbinds. Notice how both events could, in principle, add/remove multiple bridges to/from the set of CBs. For instance, in an event of type $a$, the new bound receptor could prevent some of the existing CBs from touching the perimeter. The rate at which an event of type $a$ happens is given by $k_a=\kon^T  A_\mathrm{out}/A_\mathrm{tot}$. As defined before, $\kon^T$ is the rate of forming any bridge, contained or not by the CBs. 
Events of type $b$ happen with a rate $k_b=\koff n_\mathrm{cb}$. 
$\koff$ is the rate at which a single bridge breaks (Fig.~\ref{Fig:LR}c). Instead, $\kon^T$ reads as $\kon^T=\kon \cdot n_\ell \cdot n_r$, where $\kon$ is the rate at which a bridge form from a given ligand--receptor pair (Fig.~\ref{Fig:LR}c) and $n_\ell$ ($n_r$) is the number of free ligands (receptors covered by the disk).
If the number of bound receptors $n_b$ is  steady ($n_b=\langle n_b \rangle$), the rates of forming and destroying a bridge are equal ($\kon^T  = \koff \langle n_b \rangle $). This equality allows expressing the average values of $k_a$ and $k_b$ as a function of $\koff$ and geometric factors
\begin{eqnarray}
 \langle k_a \rangle =\koff  \langle n_b\rangle { \langle A_\mathrm{out}\rangle  \over A_\mathrm{tot}} &\qquad&  \langle k_b \rangle=\koff  \langle n_\mathrm{cb}\rangle \,\, ,
\label{Eq:tauCB}
\end{eqnarray}
where $A_\mathrm{tot}=\pi R^2$ and we neglect correlations between fluctuations in $n_b$ and $A_\mathrm{out}$. The average rate at which sets of CBs are reconfigured is then given by  $\langle \tau_\mathrm{cp}\rangle^{-1}=\langle k_a \rangle +\langle k_b\rangle$  (where we treat events $a$ and $b$ as independent). Notice that events of type $a$ are the reverse of events of type $b$. These two types of events decrease and increase $A_\mathrm{out}$ by the same quantity.  Therefore, in steady conditions, we have $\langle k_a \rangle=\langle k_b \rangle$. From Eq.~\ref{Eq:tauCB} we then derive 
\begin{eqnarray}
\langle A_\mathrm{out} \rangle = \pi R^2 {\langle n_\mathrm{cb} \rangle \over \langle n_b \rangle} & \qquad \quad & \langle \tau_\mathrm{cb} \rangle={1\over 2 k_\mathrm{off} \langle n_\mathrm{cb} \rangle} \, .
\label{Eq:AoutTau}
\end{eqnarray}
In Fig.~\ref{Fig:AoutOmega}, we sample $\langle A_\mathrm{out}\rangle$ using the samples generated in Fig.~\ref{Fig:ncb}.
Simulation results agree with the theoretical predictions (Eq.~\ref{Eq:AoutTau}).
In Fig.~\ref{Fig:AoutOmega} we also sample the average area ($\langle |\Omega|\rangle $) of the configurational space $\Omega$ (Fig.~\ref{Fig:Model}c). Given that the distance between the outer-most bridges and the disk's border scales like $1/\langle n_b \rangle $, we expect that the disk could move along a given direction by a displacement $\delta\ell$, $\delta \ell \sim 1/\langle n_b\rangle $. This suggests that the configurational area scales like $\langle |\Omega|\rangle \sim \langle \delta \ell \rangle^2 \sim \langle n_b \rangle^{-2}$ in the large $\langle n_b \rangle$ limit. By fitting the simulation results of Fig.~\ref{Fig:AoutOmega}, we find that $\langle |\Omega|\rangle $ is well approximated by $\langle A_\mathrm{out}\rangle /\langle n_b \rangle$
\begin{eqnarray}
\langle |\Omega| \rangle = \pi R^2 {\langle n_\mathrm{cb} \rangle \over \langle n_b \rangle^2}
\, \, .
\label{Eq:Omega}
\end{eqnarray}


\begin{figure}[h!]
\includegraphics[width=230pt]{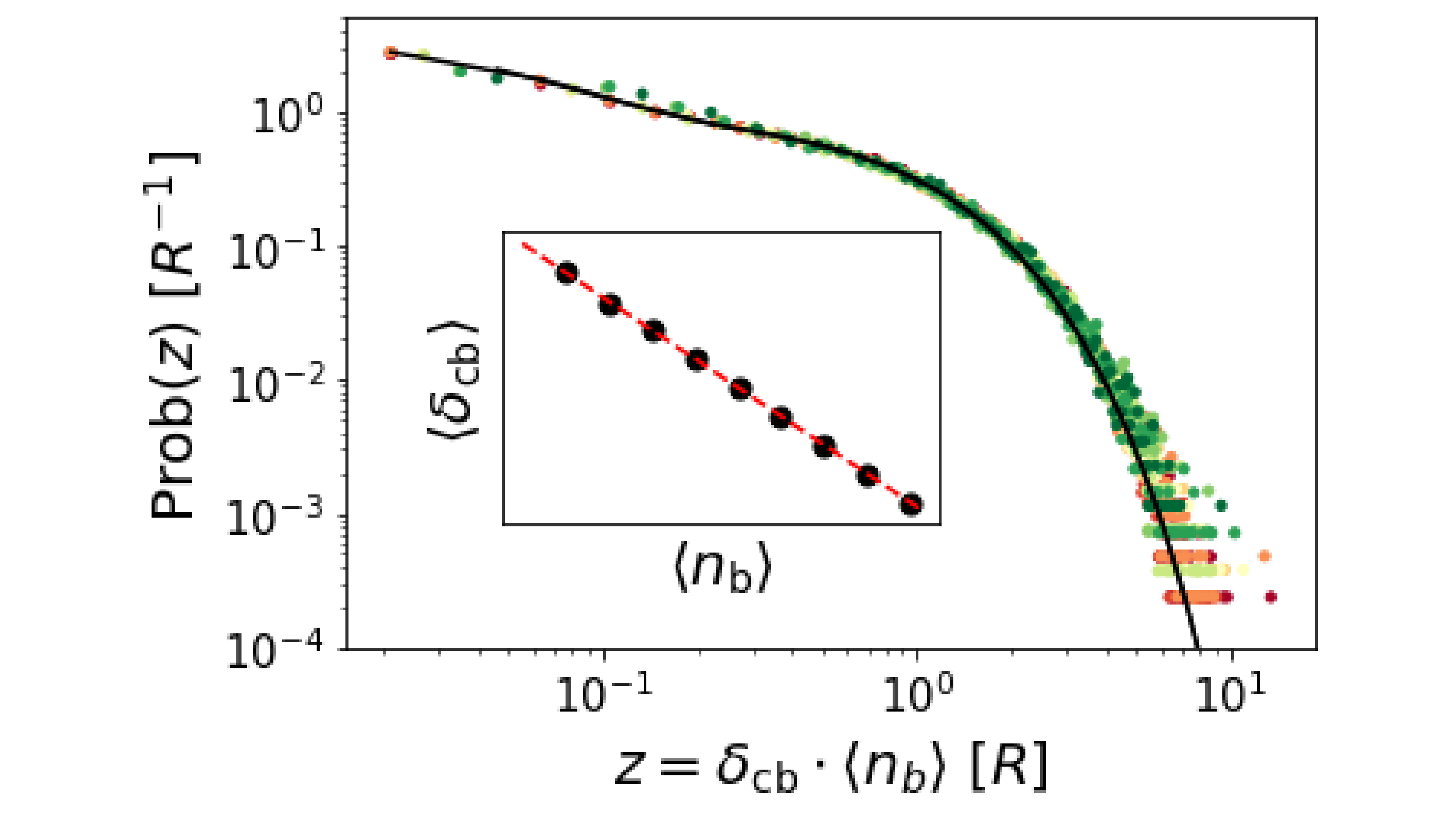}
\caption{Distribution of the displacement, $\delta_\mathrm{cb}$, of the particle following an update of the set of CBs calculated as described in the text. Symbols correspond to the set of simulations presented in Fig.~\ref{Fig:P_t_cb}. The black line is a double exponential function, $a_1\exp(-b_1\cdot z) + a_2\exp(-b_2\cdot z)$, with $a_1=0.9964$, $b_1=1.1707$, $a_2=2.67$, and $b_2=18.67$. }\label{Fig:delta}
\end{figure}

\subsection{Single-step displacement ($\delta_\mathrm{cb}$)}\label{Sec:delta}

We now estimate the average displacement of the center of mass of the particle following a change in the set of CBs ($\delta_\mathrm{cb}$) by means of numerical simulations. For a given set of CBs, we sample the position of the disk center, ${\bf x}(\Omega)$, by generating points inside $\Omega$ as done in Fig.~\ref{Fig:Model}c. While exploring $\Omega$, we attempt either to add or remove bridges. 
The probability of adding and removing a bridge are given in SI Sec.~1.
Once a reaction move changes the set of CBs, we estimate the average position of the disk for the new set of CBs, ${\bf x}(\Omega')$, and calculate $\delta_\mathrm{cb}=|{\bf x}(\Omega')-{\bf x}(\Omega)|$. By repeating the procedure, we obtain the distribution of $\tau_\mathrm{cb}$ (Fig.~\ref{Fig:P_t_cb}) and of $\delta_\mathrm{cb}$ (Fig.~\ref{Fig:delta}).

In Figs.~\ref{Fig:P_t_cb} \& \ref{Fig:delta} symbols refer to systems with an average number of bridges ranging from $\langle n_b\rangle=5$ to $\langle n_b\rangle=5120$ (as in Figs.~\ref{Fig:ncb}, \ref{Fig:AoutOmega}). Fig.~\ref{Fig:P_t_cb} shows how $\tau_\mathrm{cb}$ follows a Poisson distribution consistent with Eq.~\ref{Eq:AoutTau}. In particular, $\langle \tau_\mathrm{cb} \rangle$ does not depend on $\langle n_b \rangle$.

Concerning the distributions of $\delta_\mathrm{cb}$, Fig.~\ref{Fig:delta} shows how the distributions for different $\langle n_b \rangle$ fall on the same master curve when plotted as a function of $z=\langle n_b\rangle \cdot \delta_\mathrm{cb}$.
As a consequence $\langle \delta_\mathrm{cb} \rangle$ scales like $1/\langle n_b\rangle$ with a prefactor which is well-approximated by $R \sqrt{\pi/ \langle n_\mathrm{cb}\rangle }$ (inset of Fig.~\ref{Fig:delta})
\begin{eqnarray}
\langle \delta_\mathrm{cb} \rangle = \sqrt{\pi\over \langle n_\mathrm{cb} \rangle  } {R\over  \langle n_b \rangle } \, \, .
\label{Eq:dcb}
\end{eqnarray}
Notice that, in our calculation of $\langle \delta_\mathrm{cb} \rangle$, bridges form with uniform probability inside the particle's perimeter. The simulation setting employed in Secs.~\ref{Sec:Simulations} \& \ref{Sec:fixed} will consider explicit positions of the receptors.

\subsection{Emerging diffusion constant ($D$)}\label{Sec:D}

Having estimated $\tau_\mathrm{cb}$ and $\delta_\mathrm{cb}$, we are now in a position to predict the emerging diffusion constant $D$. The dynamics is limited by the reaction rates if $D_0 \langle \tau_\mathrm{cb} \rangle \geq \langle |\Omega| \rangle$. This inequality certainly holds in the large $\langle n_b \rangle$ limit where $\langle |\Omega| \rangle \sim \langle n_b\rangle^{-2}$ (Eq.~\ref{Eq:Omega}) and $\langle \tau_\mathrm{cb} \rangle \sim 1$ (Eq.~\ref{Eq:AoutTau}).
In a time interval $\Delta T$, the average displacement of the particle will be (neglecting correlations between sequential values of $\delta_\mathrm{cb}$)
\begin{eqnarray}
\langle \Delta {\bf x}^2 \rangle = { \Delta T \over \langle \tau_\mathrm{cp} \rangle }  \langle\delta_\mathrm{cb}\rangle^2  = 2 D \Delta T
\end{eqnarray}
from which we extract 
\begin{eqnarray}
\boxed{
D= {k_\mathrm{off} \pi R^2 \over  \langle n_b\rangle^2} \,\, .
}
\label{Eq:D}
\end{eqnarray}
For a molecular walker (Fig.~\ref{Fig:LR}b), Ref.~\onlinecite{Parker2005} recently reported the following expression for $D$
\begin{eqnarray}
D={1\over 2} {\kon \koff \over \kon + \koff} {R^2 \over \langle n_b\rangle^2} \, ,
\label{Eq:DK}
\end{eqnarray}
where $R$ is now the length of the walker's branches. This contribution identified the time-step ($\langle \tau\rangle $) of the coarse-grained dynamics with the time taken to a single ligand to bind and sequentially unbind a receptor ($\langle \tau\rangle=\kon^{-1} + \koff^{-1}$, when neglecting avidity terms in the binding event). Ref.~\onlinecite{Parker2005} then derived Eq.~\ref{Eq:DK} by fitting simulation results (with $\langle n_b\rangle<16$) with $D\sim\langle\delta\rangle^2/\langle\tau\rangle$ and $\langle \delta\rangle^2\sim \langle n_b\rangle^{-2}$ (guessed from a relation similar to Eq.~\ref{Eq:Omega}). As compared to Ref.~\onlinecite{Parker2005}, we provide a microscopic definition of $\langle \tau\rangle $ based on the identification of the set of constraining bridges.

It may look surprising that in Eq.~\ref{Eq:D} $D$ increases with the radius of the disk $R$. However, Eq.~\ref{Eq:D} holds only if the unconstrained diffusion constant $D_0$ is sufficiently high to allow the particle to sample the configurational space $\Omega$ for timescales smaller or comparable with $\langle\tau_\mathrm{cb}\rangle$. Such an approximation, for a given number of bridges $n_b$, breaks down for sufficiently large values of $R$. In such a limit, $D$ will be limited by $D_0$. Secondly, when comparing disks coated with the same density of ligands, $\langle n_b\rangle$ would scale with the area of the disk resulting in a diffusion coefficient scaling as $D\sim 1/R^2$.

Eq.~\ref{Eq:D} predicts how the ligand/receptor densities affect the mobility of the system only through $\langle n_b \rangle$. This observation explains why colloids functionalized with higher densities of ligands crystallize better\cite{wang2015crystallization}. Indeed, higher coatings increase the multivalent effect and therefore the melting temperature and $\koff$ at the melting.

In Sec.~\ref{Sec:fixed}, we generalise Eq.~\ref{Eq:D} to the case of fixed ligands.

\begin{figure}[h!]
\includegraphics[width=245pt]{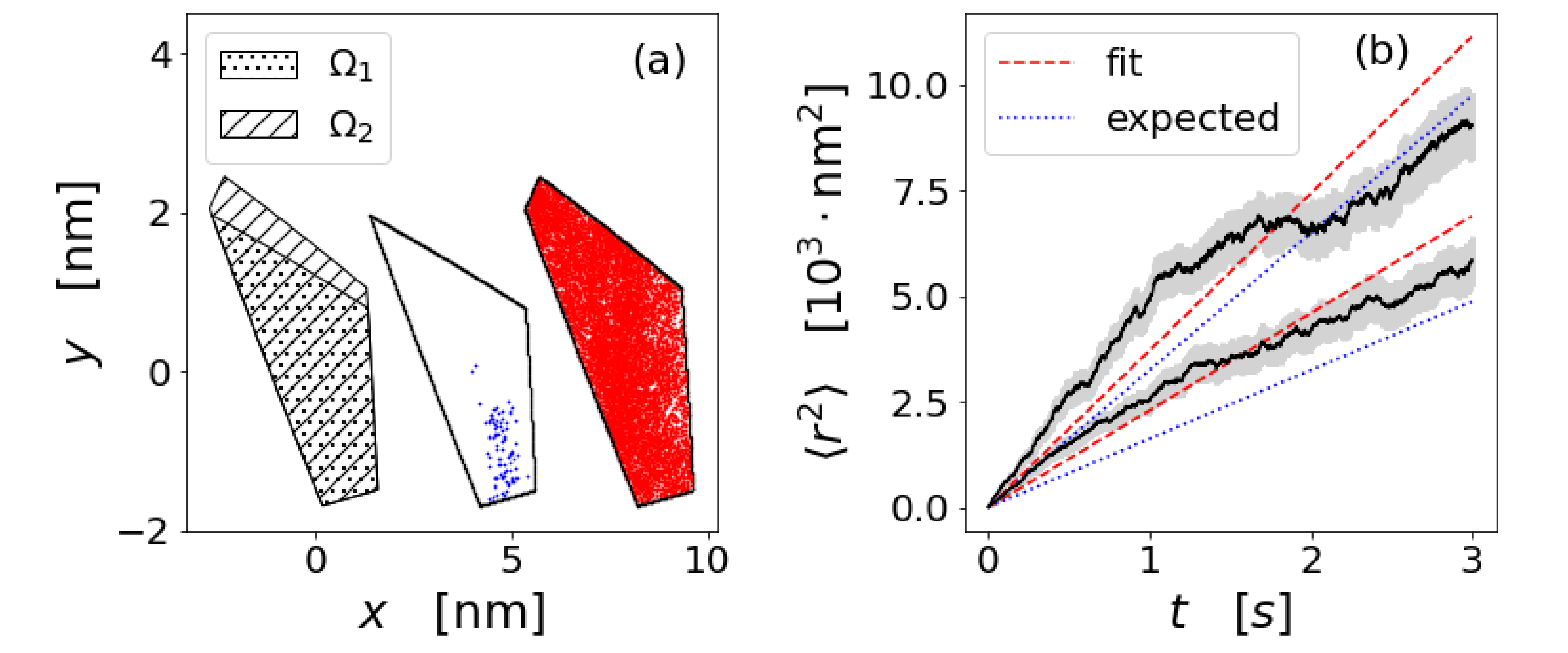}
\caption{Simulation results: mobile ligands. ({\em a})  ({\em left}) Two configurational spaces ($\Omega_i$, $i=1$, 2) visited by a $2\cdot 10^{-3}$s trajectory. ({\em center} and {\em right}) Position of the disk center (dots) while constrained to remain in $\Omega_1$ and $\Omega_2$, respectively. The disk's center of mass can explore a significant fraction of $\Omega_i$ before the latter changes as a result of an update of the CBs (Fig.~\ref{Fig:Model}). ({\em b}) Full--lines: Mean square displacements sampled  using 100 trajectories (for each curve). The shadow region represents the error bar. Dotted and dashed lines represent, respectively, a linear fitting and the theoretical curve (Eq.~\ref{Eq:D}). In ({\em b}) we use {\em i}) $\koff=636\,$s$^{-1}$, $\kon=0.318\,$s$^{-1}$ (top curve, corresponding to $L=10\,$nm, $\gamma k_{\mathrm{on},0}=10^5\,$M$^{-1}$s$^{-1}$ 
in Eq.~\ref{Eq:k_mobile}) and {\em  ii}) $\koff=318\,$s$^{-1}$, $\kon=0.159\,$s$^{-1}$  (bottom curve, corresponding to $L=10\,$nm, $\gamma k_{\mathrm{on},0}=5\cdot 10^4\,$M$^{-1}$s$^{-1}$ in Eq.~\ref{Eq:k_mobile}). In ({\em a}) we use system {\em i}). 
}\label{Fig:SimMobile}
\end{figure}

\section{Simulation validation}\label{Sec:Simulations}
Following Ref.~\onlinecite{jana2019translational}, we employ simulations merging Brownian diffusion and reaction dynamics to validate Eq.~\ref{Eq:D}. At each simulation step, we first employ the Gillespie algorithm \cite{gillespie1977exact} to update the current set of bridges from time $t$ to $t+\Delta t$, where $\Delta t$ is the simulation time step. The list of possible reactions include the breakage of a bridge and the binding of a free receptor to a free ligand. Consistently with the fact that we do not track the explicit position of the ligands (Sec.~\ref{Sec:Model}), the rate of forming a bridge is the same for all ligand--receptor pairs (well--stirred approximation). Details are reported in SI Sec.~III.A. We then attempt to evolve the center of mass of the particle, ${\bf x}_\mathrm{CM}$, using a Brownian dynamics update
\begin{eqnarray}
{\bf x}_\mathrm{CM}(t+\Delta t) = {\bf x}_\mathrm{CM}(t) + \sqrt{2 D_0 \Delta t} U \, \, ,
\label{Eq:BU}
\end{eqnarray}
where $D_0$ is the diffusion constant of the particle at the surface without bridges and $U$ is a normally distributed stochastic variable. If a bridge exits the area covered by the particle, the new configuration is rejected. High rejection rates may artificially slow down the dynamics. Therefore, as compared to previous investigations \cite{jana2019translational}, we slice the diffusion step into $N_B$ consecutive Brownian updates in which we use Eq.~\ref{Eq:BU} by replacing $\Delta t$ with $\Delta t'=\Delta t/N_B$. To limit the number of rejets, we use the results of the previous section to gauge $N_B$ as a function of $\Delta t$, $D_0$, and $\langle\delta_\mathrm{cb}\rangle$. Understanding the impact of different reaction--diffusion algorithms \cite{schoneberg2013readdy} on the mobility of the particle deserves future investigations.
\\
We consider a disk of radius $R=100\,$nm placed at the center of a square with side equal to $1\,\mu$m. We cover the surface with $N_R=17777$ randomly distributed receptors corresponding to an average receptor--receptor distance equal to $d=7.5\,$nm. This is a typical coverage density usually employed, for instance, in experiments with ligands/receptors made of DNA oligomers \cite{mognetti2019programmable,Linnee2106036118}. We do not expect that using regularly distributed receptors would have a major impact on the results of this section \cite{Parker2005}. The surface of the disk is covered with $N_L=558$ mobile ligands, matching the receptor density. The {\em on}-- and {\em off}--rates, for flexible ligands, can be calculated as \cite{parolini2016controlling,mognetti2019programmable}
\begin{eqnarray}
k_\mathrm{on}={\gamma k_\mathrm{on,0} \over \pi R^2 L}
&\qquad&
k_\mathrm{off}={ \gamma k_\mathrm{on,0}\over K_\mathrm{eq}}= \gamma k_\mathrm{on,0} K_D \, ,
\label{Eq:k_mobile}
\end{eqnarray}
where $K_\mathrm{eq}$, $K_D$, and $k_{\mathrm{on},0}$ are, respectively, the equilibrium costant, the dissociation constant, and the {\em on} rate measured for ligands and receptors free in solution (i.e.\ not anchored to the particle and surface). $L$ is a length comparable with the size of the ligands. $\gamma$ ($\gamma<1$) is a non-dimensional term accounting for the fact that the diffusion constants of anchored ligands/receptors are reduced as a result, e.g., of the drag exherted by the bilayer onto the anchoring point \cite{evans_sackmann_1988,merminod2021avidity,Linnee2106036118} or by the solvent onto the polymeric backbone supporting the reactive complex \cite{sandholtz2019physical}. $k_{\mathrm{on},0}$ can span multiple orders of magnitudes (e.g., Refs.~\onlinecite{peck2015rapid,reiter2019force,Delgusteeaat1273}) and is always smaller than the diffusion--limited value $k^{(\mathrm{dl})}_{\mathrm{on},0}\approx 10^8\,M^{-1}s^{-1}$ (where $M$ is the molarity unit, $1\cdot M=0.602\,$nm$^{-3}$), calculated using the Smoluchowski's equation for a nanometer--size probe. Importantly, our simulations model bridge formation/denaturation as Poissonian events and therefore neglect rebinding events \cite{van2005green}. This is not expected to be a serious drawback as the emergent diffusion is order of magnitudes smaller than the diffusions constant of the ligands (which could be $\approx \mu m^2/s$). The calculation of the {\em on}/{\em off} rates of ligands tethered to a surface is currently debated \cite{xu2015binding}.

Fig.~\ref{Fig:SimMobile} reports our simulation results for the mobile ligand system with $\langle n_b \rangle= 111$. Fig.~\ref{Fig:SimMobile}a (left) shows the two configurational spaces ($\Omega_1$ and $\Omega_2$) sampled by an $\approx 2\cdot 10^{-3}$s  trajectory. Figs.~\ref{Fig:SimMobile}a (center and right) show the center of mass trajectory (points) while constrained to remain in $\Omega_1$ and $\Omega_2$, respectively. 
Notice that the trajectory exploring $\Omega_1$  is much shorter than the one constrained by $\Omega_2$. The variability of the trajectory length is related to the stochastic nature of $\tau_\mathrm{cb}$. Importantly, in both cases, the trajectories can travel a distance comparable to the size of $\Omega_i$. This justifies the assumptions employed in the previous section and clarifies how the long-time diffusion is controlled by the evolution of $\Omega$ arising from the formation/rupture of bridges.

In Fig.~\ref{Fig:SimMobile}b, we report the mean squared displacement for two pairs of {\em on} and {\em off} rates with $\kon/\koff$ constant (resulting in the same $\langle n_b \rangle$). In both cases, we show how Eq.~\ref{Eq:D} predictions agree with simulation results. A validation employing different $\langle n_b \rangle$ is presented in the next section. The simulation programs employed to obtain Fig.~\ref{Fig:SimMobile}b is available online \footnote{URL: \url{https://github.com/bmognetti/SlidingDiffusionConstant_Mobile.git}}.

\begin{figure}[htb]
\includegraphics[width=250pt]{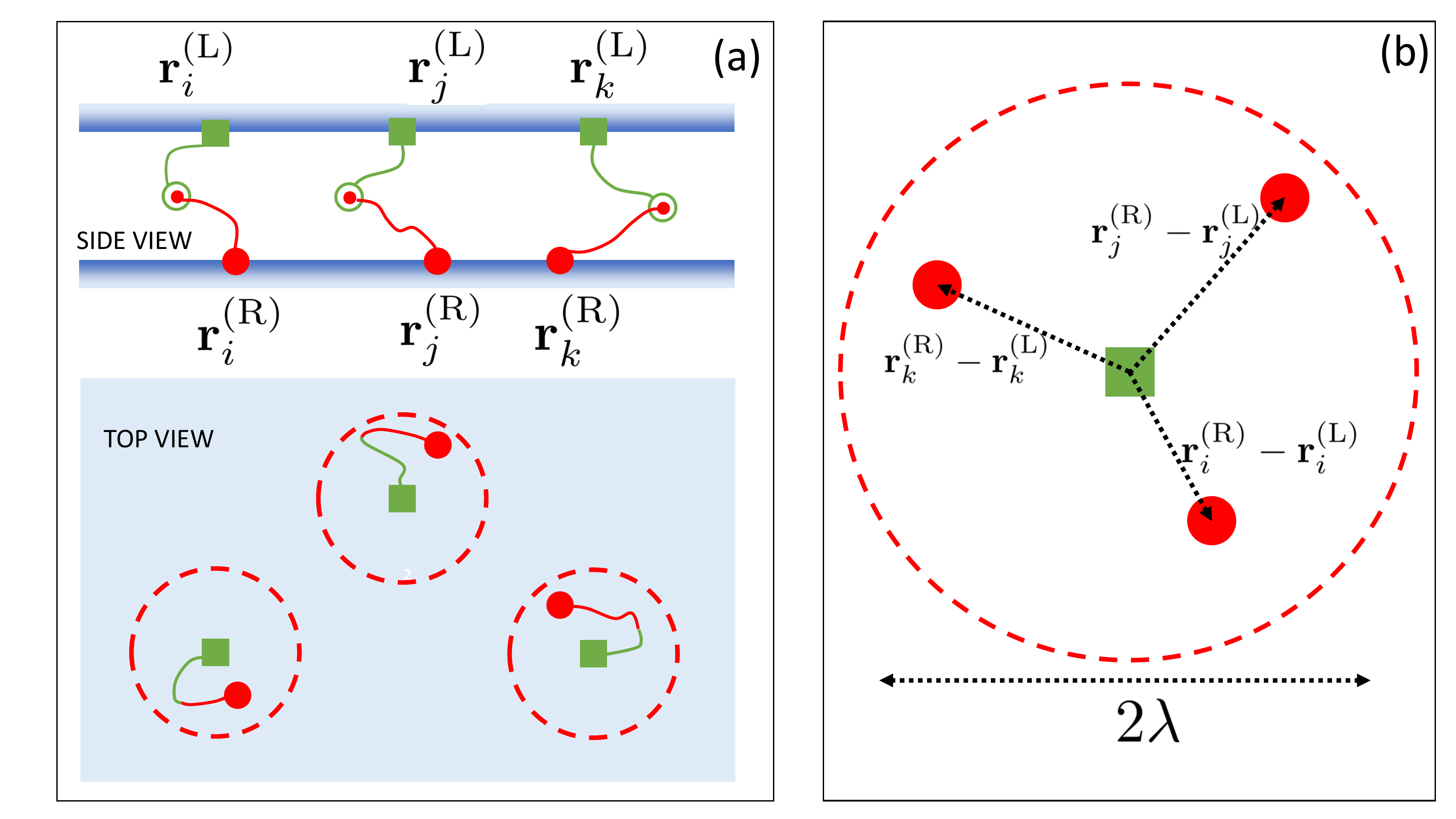}
\caption{Particles with fixed ligands. ({\em a}) A ligand--receptor pair can form a bridge if the distance between the tethering points (dots and squares) is smaller than $\lambda$. We report only ligands/receptors forming a bridge.  ({\em b}) By representing the relative position of the bound receptors as compared to the conjugated ligands, fixed ligand systems are mapped into mobile ligand systems with $R=\lambda$ (Fig.~\ref{Fig:LR}).  }\label{Fig:Fixed_Model}
\end{figure}

\begin{figure*}[htb]
\includegraphics[width=500pt]{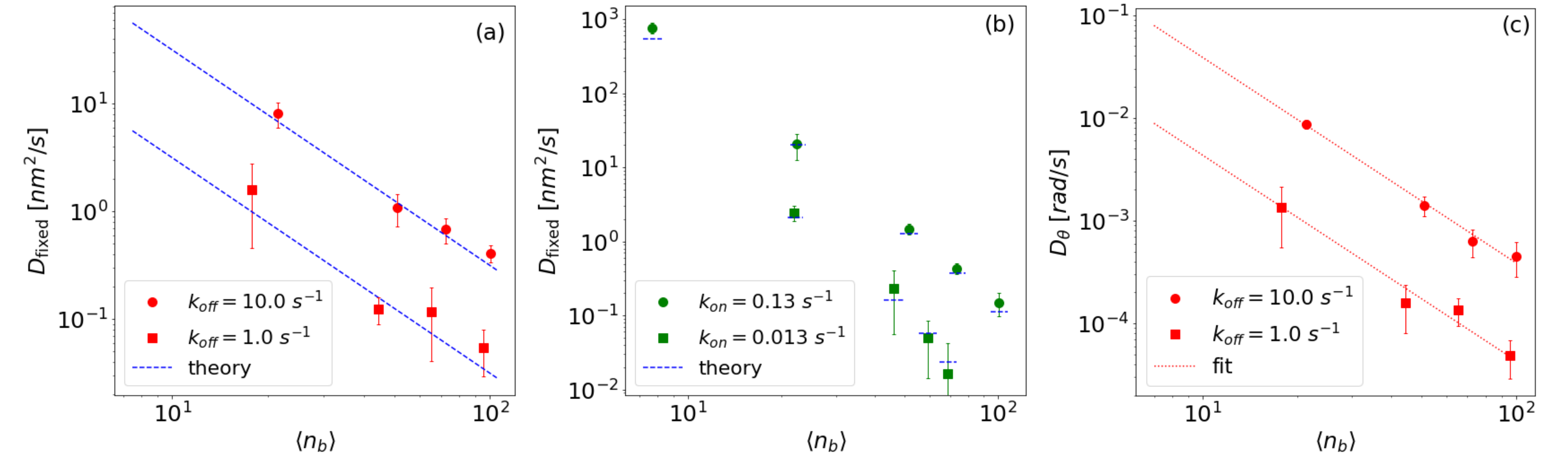}
\caption{Simulation results: fixed ligands. ({\em a} and {\em b}) Translational diffusion constant at a given  $\koff$ and $\kon$, respectively, and different numbers of bridges (as obtained by changing, respectively, $\kon$ and $\koff$). ({\em c}) Rotational diffusion constant at a given $\koff$ and different $\langle n_b\rangle$.
The lines in ({\em a}) and ({\em b}) are the theoretical predictions (Eq.~\ref{Eq:Dfixed}) while in ({\em c}) are the fits to the data points. We generate 100, $1\,$s trajectories for each data point. We calculate the error bars using 5 independent estimates of the diffusion constant obtained by dividing the batch of trajectories into 5 subgroups. The error bars on the number of bridges are smaller than the symbol.}\label{Fig:Dfixed}
\end{figure*}

\section{Fixed ligands}\label{Sec:fixed}

We now discuss the case in which ligands are not mobile but fixed to the surface of the particle. We define by ${\bf r}^\mathrm{(L)}_i$ the coordinates of the ligands' tethering points (square symbols in Fig.~\ref{Fig:Fixed_Model}a). A receptor, grafted in ${\bf r}^\mathrm{(R)}_i$, can then form a bridge with ligand $i$ if $|{\bf r}^\mathrm{(R)}_i-{\bf r}^\mathrm{(L)}_i|<\lambda$, where $\lambda$ is the maximal lateral distance between the tethering points of a ligand--receptor pair forming a bridge. For a given set of bridges, the possible configurations of the particle are then the ones in which none bound receptor exits the circle of radius $\lambda$ centered over the conjugated ligand (dashed circles in Fig.~\ref{Fig:Fixed_Model}a).
By shifting all the bound receptor's position by the position of the corresponding conjugated ligand (${\bf r}^\mathrm{(R)}_i \to {\bf r}^\mathrm{(R)}_i - {\bf r}^\mathrm{(L)}_i$), as done in Fig.~\ref{Fig:Fixed_Model}b, we show how the possible configurations of the particle (for a given orientation) correspond to the possible configurations of a disk of radius $\lambda$ with mobile ligands binding uniformly distributed receptors (Fig.~\ref{Fig:Model}). In particular, the area available to the center of the particle with fixed ligands (for a given orientation) will follow from the expression of $\langle|\Omega|\rangle$ (Eq.~\ref{Eq:Omega}) by replacing $R$ with $\lambda$. Similarly,  $\lambda$ replaces $R$ in the average displacement $\langle \delta_\mathrm{cb}\rangle$ (Eq.~\ref{Eq:dcb}), while the average time to reconfigure a set of constraining bridges $\langle \tau_\mathrm{cb}\rangle$ (Eq.~\ref{Eq:AoutTau}) does not change. Finally, the sliding diffusion constant will mimic Eq.~\ref{Eq:D} as follows
\begin{eqnarray}
\boxed{
D_\mathrm{fixed}={k_\mathrm{off} \pi\lambda^2 \over \langle n_b\rangle^2} \, .
}
\label{Eq:Dfixed}
\end{eqnarray}
The previous equation implies that, if $\lambda \ll R$, the emerging diffusion for fixed ligands will be much smaller than for mobile ligands. Moreover, the configurational area available to the particle's center of mass will also decrease. The latter observation implies that the assumptions underlying Eq.~\ref{Eq:D} are less severe in the present case: the trajectory will have the same time ($\tau_\mathrm{cb}$) to explore configurational spaces ($\Omega_i$) smaller than the ones considered in Fig.~\ref{Fig:SimMobile}a.

To validate Eq.~\ref{Eq:Dfixed}, we simulate a $d=$100$\,$nm diameter disk carrying  $N_L=200$ randomly distributed ligands and a surface featuring 0.1 receptors$/$nm$^{-2}$. As for mobile ligands, we do not expect that major differences would arise from using surfaces with regular patterns of ligands/receptors (except for settings in which bridges could form only in particular alignment conditions). Ligands can bind receptors placed at a lateral distance smaller than $\lambda = 10.0\,$nm (Fig.~\ref{Fig:Fixed_Model}). For fixed ligands, the relations between $\kon$, $\koff$ and $k_{\mathrm{on},0}$, $k_{\mathrm{off},0}$ are different from what stated by Eq.~\ref{Eq:k_mobile}. In general, $\kon$ and/or $\koff$ are also a function of the relative distance between the tethering points (e.g.~Ref.~\onlinecite{mognetti2019programmable}). In this section, consistent with the fact that bridges do not exert any force onto the particle when the distance between tethering points is smaller than $\lambda$, we consider configuration--independent rates. This choice is consistent with previous literature (e.g.~\onlinecite{vahey2019influenza}). In Sec.~\ref{Sec:Discussion}, we discuss how different bond potentials between bridged tethering points could affect our findings.

Given that ligands are fixed, the particle's orientation, $\theta$, is a non--degenerate variable. In particular, the system features an emerging rotational diffusion constant, $D_\theta$.
The (rotational) diffusion constant of the free particle is $D_0=10^5\,$nm$^2$/s ($D_{\theta,0}=100\,$rad/s). 
As found in Sec.~\ref{Sec:Simulations}, in the limit of validity of the theory, $D_0$ and $D_{\theta,0}$ do not affect the emerging diffusion constants ($D_\mathrm{fixed}$ and $D_\theta$).
In all simulations, we use $\Delta t=10^{-5}\,$s and $N_B = 10$ (Sec.~\ref{Sec:Simulations}).

Figs.~\ref{Fig:Dfixed}a and \ref{Fig:Dfixed}b report simulation results for $D_\mathrm{fixed}$ as a function of $\langle n_b\rangle$. We consider sets of runs with $\koff$ constant and different $\kon$ (Fig.~\ref{Fig:Dfixed}a) and sets with $\kon$ constant and different $\koff$ (Fig.~\ref{Fig:Dfixed}b). The resulting average number of bridges ranges between $\langle n_b\rangle \approx 8$ and $\langle n_b\rangle \approx 100$. Systems with $\koff$ constant can be directly compared with Eq.~\ref{Eq:Dfixed} (dashed lines in Fig.~\ref{Fig:Dfixed}a). For the simulations at constant $\kon$, the predictions of Eq.~\ref{Eq:Dfixed} are given by the horizontal dashed lines calculated using the different $\koff$ employed by the simulations. In the large $N_L$ limit (at constant $\langle n_b\rangle$), it can be shown that, for constant $\kon$, $D_\mathrm{fixed}$ decreases as $1/\langle n_b\rangle^3$ (SI Sec.~IV). Overall, the simulation results of Fig.~\ref{Fig:Dfixed}a are in excellent agreement with Eq.~\ref{Eq:Dfixed} for all values of $\kon$ and $\koff$. $D_\mathrm{fixed}$ is not a function of the particle's shape, as highlighted by the geometric construction in Fig.~\ref{Fig:Fixed_Model} and SI Fig.~3a. In particular, contrarily to $D_0$, $D_\mathrm{fixed}$ is always isotropic.

From the simulation results, it is not immediate to speculate about systematic errors in Eqs.~\ref{Eq:D} and \ref{Eq:Dfixed}. Our theory neglects that, for low values of $\langle n_b\rangle$, the emerging diffusion constant could be limited by $D_0$. However, in Figs.~\ref{Fig:SimMobile}b, \ref{Fig:Dfixed}a,  and \ref{Fig:Dfixed}b, the theory overestimates simulations rarely. Possible systematic errors could arise from treating successive unit displacements $\delta_\mathrm{cb}$ as independent variables while some degree of correlation is expected given that successive $\Omega_i$ share a subset of constraining bridges. A detailed study of the statistical properties of the coarse-grained dynamics underlying our model deserves future investigations.

Fig.~\ref{Fig:Dfixed}c and SI Fig.~2 report the results for the rotational diffusion constant, $D_\theta$  {\em vs} $n_b$, at constant $\koff$ (Fig.~\ref{Fig:Dfixed}c) and $\kon$ (SI Fig.~2). $D_\theta$ mimics the trends observed for $D_\mathrm{fixed}$. By fitting the data points at constant $\koff$ (Fig.~\ref{Fig:Dfixed}c) we find 
\begin{eqnarray}
D_\mathrm{\theta} = c {\koff \pi  \over \langle n_b\rangle^2 }  
\end{eqnarray} 
with $c\approx 0.13$. Contrarily to what observed for $D_\mathrm{fixed}$, $D_\theta$ is affected by the shape of the particle. In particular, $D_\theta$ decreases when considering more elongated particles (SI Fig.~3b). The simulation programs employed to obtain Fig.~\ref{Fig:Dfixed} is available online \footnote{URL: \url{https://github.com/StevensLaurie/Disk_Diffusion}}.

\section{Discussions and conclusions}\label{Sec:Discussion}

In  this paper we have derived two simple equations (Eqs.~\ref{Eq:D}, \ref{Eq:Dfixed}) for the emergent sliding diffusion constant of particles bound to a surface through reversible linkages. We validated the theoretical predictions by employing reaction diffusion simulations. The diffusion constant is  controlled by the rate at which ligand--receptor pairs unbind ($\koff$), the average number of bridges ($\langle n_b\rangle$), and the area of the surface accessible to a single ligand ($\pi R^2$ in Eq.~\ref{Eq:D} or $\pi \lambda^2$ in Eq.~\ref{Eq:Dfixed}). Our results strictly apply to the case of fixed receptors. For mobile receptors, particles anchored to a surface through static bridges can laterally diffuse\cite{merminod2021avidity,evans_sackmann_1988}.

Ref.~\onlinecite{marbach2021nanocaterpillar} recently presented numerical and analytic predictions of the emerging sliding diffusion of a 1D system. As compared to our setting, this contribution employs harmonic (rather than square-well) bond potentials $h$ (Fig.~\ref{Fig:LR}c) to model the forces exerted by the bridges onto the particle. The prediction of Ref.~\onlinecite{marbach2021nanocaterpillar} for $D$ (in the large $\langle n_b\rangle$ limit) reads as follows 
\begin{eqnarray}
D={k_B T \over \Gamma + \langle n_b\rangle  \gamma +  \langle n_b\rangle {k \over \koff } +  \langle n_b\rangle  {\gamma \kon \over \koff } } \, ,
\label{Eq:DNY}
\end{eqnarray}
where $\Gamma$ and $\gamma$ are the friction coefficients controlling the diffusion of the free particle ($D_0=k_B T / \Gamma$) and of the reactive tips of the ligands, while $k$ is the spring constant of the bond potential. Ref.~\onlinecite{marbach2021nanocaterpillar} predicts how, for $\langle n_b\rangle\to 0$, the emerging sliding diffusion is limited by $D_0$. In the many--bridge limit, it predicts a $1/\langle n_b\rangle $ scaling law which is in contrast with Eqs.~\ref{Eq:D} \& \ref{Eq:Dfixed}. Such a disagreement unlikely arises from the fact that Ref.~\onlinecite{marbach2021nanocaterpillar} employs a 1D system. Indeed, adapting our model to a 1D system would not impact the $1/\langle n_b\rangle^2$ scaling found in Eqs.~\ref{Eq:D}, \ref{Eq:Dfixed}. It sounds more reasonable that the two different types of bond potentials could play a crucial role. This observation points to the fact that polymeric details of the ligand/receptor molecules, not directly linked to the rate coefficients ($k_{\mathrm{on},0}$ and $\koff$), may have a major impact on the mobility of the particle. The bond potential is largely tunable. For instance, ligands resembling fully flexible ideal polymers or thin rods would result, respectively, in a quadratic and logarithmic bond potential \cite{Treloar1946}.

Existing literature in biophysics, employing models with cross-linked objects, already reported emerging friction terms $\zeta$ ($\zeta \sim 1/D$) which are non--linear in the number of crosslinkers. For instance, Wierenga {\em et al.}~\cite{wierenga2020diffusible} show how the friction between two sliding cytoskeletal filaments is super exponential in the number of crosslinkers, $\langle n_b\rangle$. In this system, the highly non-linear increase of $\zeta$ in $\langle n_b\rangle $ arises from the necessity of having a specific sequence of $\langle n_b\rangle$ unbinding-rebinding events to observe a net ($O(1)$) relative displacement of the filaments. Intriguingly, cooperative effects may also reduce the friction constant. For instance, Fogelson {\em et al.}~\cite{fogelson2019transport} study monovalent particles diffusing over a substrate decorated by receptors. In this system, multiple particles diffuse faster (as compared to a diluted system) as a result of an effective repulsive force mediated by the functionalized surface.

When comparing the predictions of Eqs.~\ref{Eq:D}, \ref{Eq:Dfixed} with experiments employing spherical particles, one should consider that the latter tend to roll rather than slide \cite{C8SM01430B,jana2019translational}. The prediction of the emergent rotational diffusion constant deserves future investigations. As compared to the present investigation, Ref.~\onlinecite{jana2019translational} studied large rates constants ($k_{\mathrm{on},0} \geq 10^6\,$M$^{-1}$s$^{-1}$). Due to the short simulation times employed (milliseconds rather than seconds as in this study), Ref.~\onlinecite{jana2019translational} did not detect sliding motion for small $k_{\mathrm{on},0}$ and large $n_b$. In the regime of validity of Eqs.~\ref{Eq:D} \& \ref{Eq:Dfixed}, we predict how the diffusion constant is not a function of the drag exerted by the medium onto the particle. This result could be tested in experiments, for instance, by adding inert polymers to increase the viscosity of the medium without affecting the biochemistry of the system.
On the other hand, sliding is likely the major component of the mobility of non-spherical particles, like the elongated form of Influenza A Virus (IAV) \cite{vahey2019influenza}. However, it is believed that IAV particles achieve their mobility because of the catalytic effect of neuramidase ligands cleaving the sialic acid present on the receptors \cite{de2020influenza}. The outcomes of the present study (as well as Refs.~\onlinecite{marbach2021nanocaterpillar,Parker2005}) will help quantitatively assess the contribution to the mobility of the particle due to the presence of activity in IAV and other biological systems. Finally, for fixed ligands, we predict that elongated particles would diffuse homogeneously. This observation will allow for the detection of catalytic activity which has been shown to be responsible for persistent motions \cite{de2020influenza}.





\section*{Supplementary Material}
See supplementary material for figures supporting the claims made in the manuscript and details about the simulation alghoritms.

\section*{Acknowledgements}

We thank  M.~Holmes-Cerfon (New-York University, USA) for bringing Refs.~\onlinecite{efron1965convex}, \onlinecite{Parker2005}, and \onlinecite{fogelson2019transport} to our attention. We thank M.\ Holmes-Cerfon (New-York University, USA), S.\ Korosec (Simon Fraser University, Canada), and N.\ R.\ Forde (Simon Fraser University, Canada) for useful discussions. JL is supported by a BAEF grant. LS is supported by a seed grant from the Interuniversity Institute of Bioinformatics in Brussels (IB2). BMM is supported by a PDR grant of the FRS-F.N.R.S.~(grant n$^\circ$T015821F). Computational resources have been provided by the Consortium des \'Equipements de Calcul Intensif (C\'ECI), funded by the F.R.S.-FNRS under Grant No.\ 2.5020.11 and by the Walloon Region.




\providecommand{\noopsort}[1]{}\providecommand{\singleletter}[1]{#1}%
%


\end{document}


\title{SUPPLEMENTARY INFORMATION: Sliding across a surface: particles with fixed and mobile ligands}
\author{Janna Lowensohn}
\thanks{These two authors contributed equally}
\affiliation{ 
Center for Nonlinear Phenomena and Complex Systems, Code Postal 231, Universit\'e Libre de Bruxelles, Boulevard du Triomphe, 1050 Brussels, Belgium
}%
\author{Laurie Stevens}
\thanks{These two authors contributed equally}
\affiliation{ 
Interuniversity Institute of Bioinformatics in Brussels, ULB-VUB, La Plaine Campus, 1050 Brussels, Belgium
}%
\affiliation{ 
Center for Nonlinear Phenomena and Complex Systems, Code Postal 231, Universit\'e Libre de Bruxelles, Boulevard du Triomphe, 1050 Brussels, Belgium
}%
\affiliation{ 
Applied Physics Research Group, Vrije Universiteit Brussel, Pleinlaan 2, 1050 Brussels, Belgium 
}
\author{Daniel Goldstein}
\affiliation{ Department of Physics and Astronomy, Tufts University,
574 Boston Avenue, Medford, Massachusetts 02155, USA
}
\author{Bortolo Matteo Mognetti}
\email{Bortolo.Matteo.Mognetti@ulb.be}
\affiliation{ 
Center for Nonlinear Phenomena and Complex Systems, Code Postal 231, Universit\'e Libre de Bruxelles, Boulevard du Triomphe, 1050 Brussels, Belgium}
\maketitle


\section{Supplementary Figures}

\begin{figure}[ht!]
\vspace{-0.7cm}
\includegraphics[width=250pt]{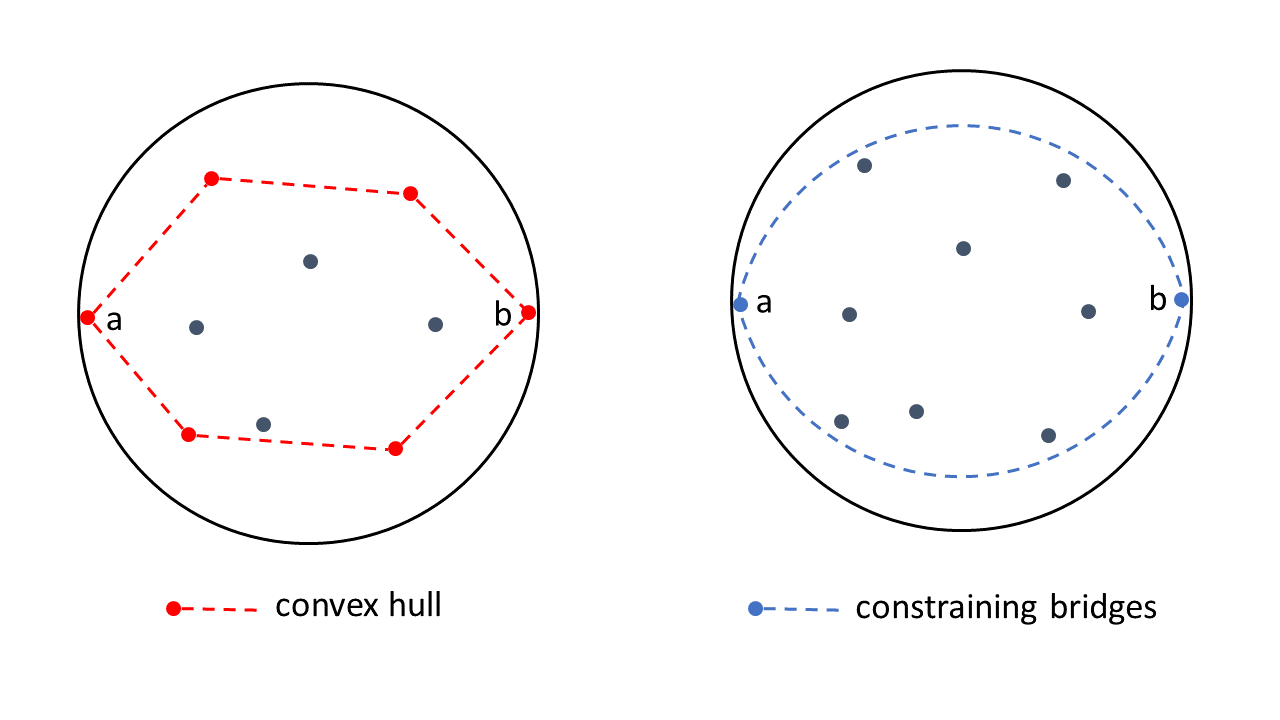} 
\vspace{-1.2cm}
\caption{ The number of constraining bridges is smaller than the number of vertices of the convex hull. {\em  (left)} The convex hull of an ensemble of points $Q$ is the smallest convex polygon (dashed line) containing all the points belonging to $Q$. {\em (right)} Two diametrically opposite points ($a$ and $b$) impede any other points from touching the border of the disc. }
 \end{figure}
\begin{figure}[ht!]
\vspace{-0.2cm}
\includegraphics[width=180pt]{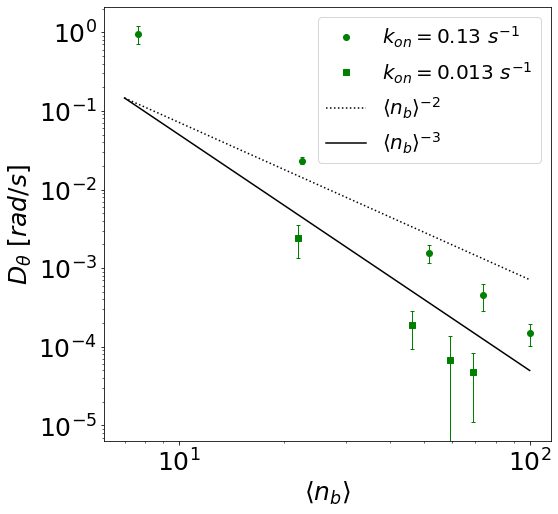} 
\caption{ Simulation results: fixed ligands.  Rotational diffusion constant at a given $\kon$ and for different numbers of bridges as obtained by changing $\koff$. The rotational diffusion constant is compatible with a $\sim \langle n_b\rangle^{-3}$ scaling law. }
 \end{figure}
\begin{figure}[ht!]
\vspace{-0.2cm}
\includegraphics[width=250pt]{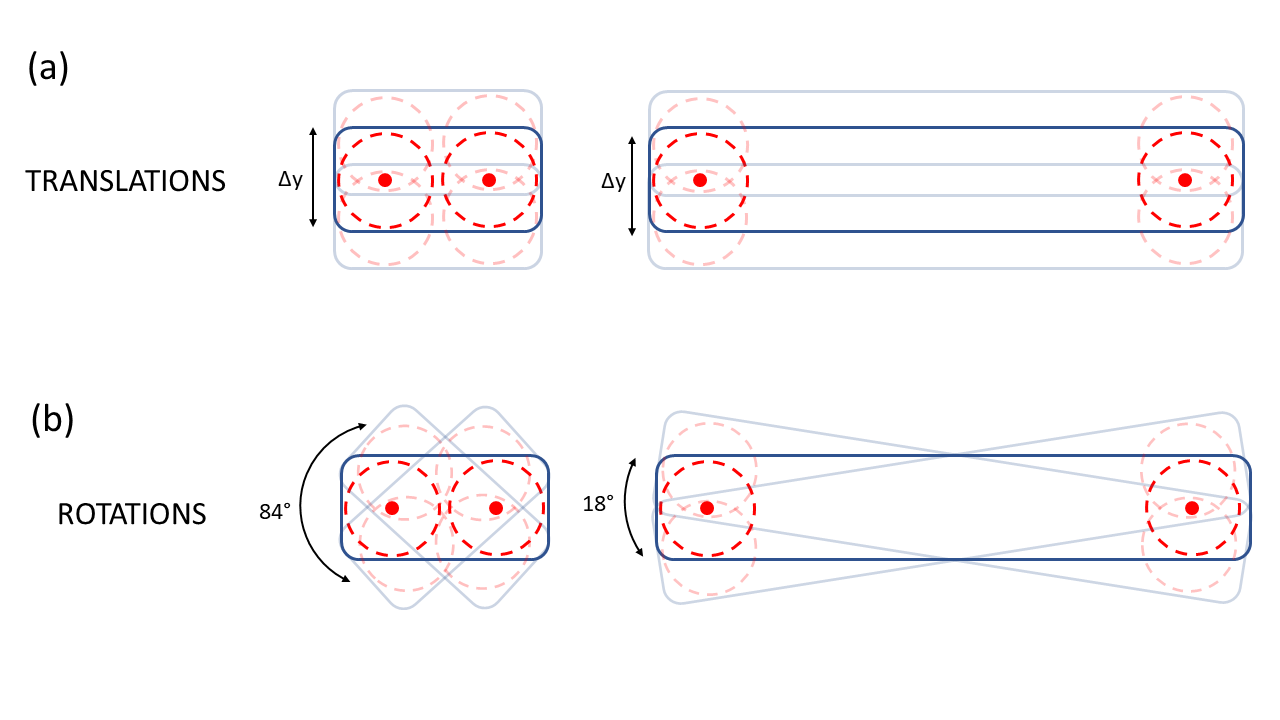} 
\caption{ In systems with fixed ligands, the particle shape affects the rotational but not the translational dynamics. We consider two objects carrying two fixed ligands (corresponding to the center of the circles of radius $\lambda$, see Main Fig.~8) bound to a fixed receptor (red dots). ({\em a}) The configurational spaces available to the center of mass of the the two objects are equal. ({\em b}) The shorter object can explore a broader set of orientations than the elongated particle.  }
 \end{figure}

\section{Monte Carlo calculation of $\delta_\mathrm{cx}$}

Bridges are added and removed at a constant rate of $\kon^T$ and  $\koff$, respectively. The average number of bridges, $\langle n_b\rangle$, then reads as follows 
\begin{eqnarray}
\langle n_b \rangle = {\kon^T \over \koff} \, . 
\end{eqnarray}
Importantly, in simulations, the number of bridges $n_b$ fluctuates around $\langle n_b\rangle$. In each cycle, we either add or remove a bridge with probability $p_a$ and $p_r$
\begin{eqnarray}
p_a &=& {\kon^T \over \kon^T+ \koff n_b} = {\langle n_b\rangle \over \langle n_b\rangle + n_b} \, \, ,
\\ 
p_r &=& {\koff n_b \over \kon^T+ \koff n_b} = { n_b \over \langle n_b \rangle+  n_b} \, \, .
\end{eqnarray}
With probability $p_a$, a bridge is created at a position uniformly distributed inside the disk. With probability $p_r$, an existing bridge is chosen uniformly from within the $n_b$ ones already present in the system. After adding/removing a bridge, we check if the set of constraining bridges (CBs) has changed. For each simulation employed in Fig.~6, we generate $5\cdot 10^4$ different sets of CBs (and therefore an equal number of displacements $\delta_\mathrm{cb}$). The time interval of each cycle is 
\begin{eqnarray}
\Delta t = {1 \over \koff n_b + \kon^T} \approx {1 \over 2 \langle n_b \rangle \koff }
\end{eqnarray}
$\tau_\mathrm{cb}$ (Main Fig.~5) is then sampled by multiplying $\Delta t$ for the number of cycles between two CB updates.

As compared to the simulations of Main Sec.~IV, the on rates ($\kon^T = \kon \cdot n_\ell \cdot n_r$) used in this section are not a function of the current number of bridges $n_b$. This would correspond to systems in which the total number of free ligands (receptors) on the disk (under the disk), $n_\ell$ ($n_r$), is not sensibly depleted by the number of bridges. Simulations employed in Main Sec.~IV also track explicitly the position of bound and unbound receptors while, in this section, the position of new bridges (corresponding to the receptor's position) are generated uniformly within the disk.

\begin{figure}[ht!]
\includegraphics[width=250pt]{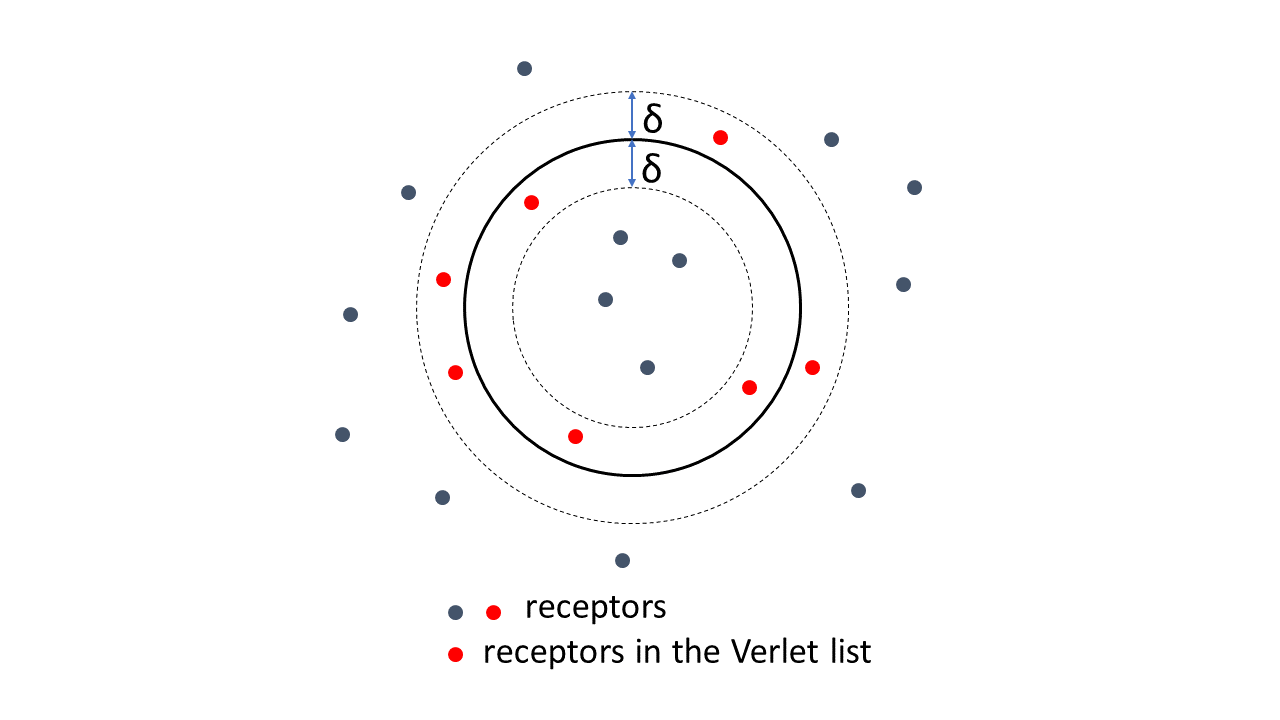} 
\caption{ The Verlet list employed for mobile ligand systems contains all receptors with  distances from the edge of the disk smaller than $\delta$. A new list is initialised once the disk diffuses a distance bigger than $\delta$.}\label{Fig:VL}
 \end{figure}

\section{Diffusion-reaction simulations}

In this section we present the simulation algorithm employed in Main Secs.~IV (mobile ligands) and V (fixed ligands). If $\Delta t$ is the simulation time step, each run consists of $N_c$ cycles,  $N_c=t_\mathrm{fin}/\Delta t$, where $t_\mathrm{fin}$ is the length of the trajectory:
\vspace{0.15 cm} \\
1 $t=0$\\
2 while$(t<t_\mathrm{fin})$:\\
3 \hspace{0.5 cm} Fire Reactions$(\Delta t)$\\
4 \hspace{0.5 cm} $i_B=0$\\
5 \hspace{0.5 cm} while$(i_B<N_B)$: \\
6 \hspace{1. cm} BrownianDiffusion$(\Delta t/N_B)$\\
7 \hspace{1. cm} $i_B+=1$ \\
8 \hspace{0.5 cm} $t+=\Delta t$ \\
\vspace{0.15cm}\\
Fire Reactions($\Delta t$) boosts the reaction dynamics by a time interval $\Delta t$ using the Gillespie algorithm. We describe in detail this routine adapted to the case of mobile and fixed ligands in the next two sections. BrownianDiffusion$(\Delta t')$ attempts a Brownian dynamics step which is accepted unless a bridge exits from the disk perimeter (for mobile ligands) or becomes overstretched (when the ligand-receptor distance becomes larger than $\lambda$, for fixed ligands).

Fire Reactions() and Brownian Diffusion() require the list of all the receptors which could potentially form a bridge. We update this list using Verlet lists. For mobile ligands, we initialise a single Verlet list including all the receptors found at a distance smaller than $\sigma$ from the edge of the disk (see Fig.~\ref{Fig:VL}). For fixed ligands, we use a different Verlet list for each ligand $i$ listing all the receptors closer than $d$ to $i$ ($d>\lambda$).

\subsection{Gillespie algorithm: mobile ligands}

As usually done in the Gillespie algorithm, we calculate the total affinity of forming ($a_\mathrm{on}^T$) or breaking ($a_\mathrm{off}^T$) a bridge as 
\begin{eqnarray}
a_\mathrm{on}^T=(N_L-n_b)(N_R-n_b) \kon 
& \qquad &
a_\mathrm{off}^T=n_b \koff
\label{affinities:mobile}
\end{eqnarray}
where $N_L$ and $N_R$ are, respectively, the total number of ligands and receptors covered by the disk. $N_R$ is a function of the disk's position and is updated using the Verlet lists described in the previous section. In view of the ligands' mobility, we use a well-stirred approximation in which a receptor can bind to all free ligands belonging to the disk with a rate constant equal to $\kon$. The Gillespie algorithm samples the next reaction to be fired (with a probability proportional to its affinity, line 5 below) along with the time for it to happen, $t_\mathrm{reac}$. $t_\mathrm{reac}$ is sampled from a Poisson distribution with average equal to $1/a^T$, $a^T=a_\mathrm{on}^T+a_\mathrm{off}^T$ (line 3 below).
The chart flow of the program for firing reactions with mobile ligands is as follows 
\vspace{0.15 cm} \\
1 $t_\mathrm{reac}=0$\\
2 while$(t_\mathrm{reac}<\Delta t)$:\\
3 \hspace{0.5 cm} $t_\mathrm{reac}+=-\log(1-u_1)/a^T$   \\
4 \hspace{0.5 cm} if$(t_\mathrm{reac}<\Delta t)$:\\
5 \hspace{1. cm} if$(u_2<a_\mathrm{on}^T/a^T)$: \\
6 \hspace{1.5 cm} bind one of the free receptors \\
7 \hspace{1.5 cm} update the affinity lists: $a_\mathrm{on}$, $a_\mathrm{off}$  \\
8 \hspace{1. cm} else: \\
9 \hspace{1.5 cm} unbind one of the bound receptors \\
10 \hspace{1.5 cm} update the affinity lists: $a_\mathrm{on}$, $a_\mathrm{off}$  \\
11 \# $u_i$: independent, uniform random numbers, $u_i\in[0,1)$
\vspace{0.15cm}\\
Notice that the last reaction which would happen at $t_\mathrm{rec}>\Delta t$ is refused (as it fails the check at line 4) and the clock is set back to $t+\Delta t$ before entering the next cycle of reactions (see previous section). This procedure does not bias the reaction dynamics given that the reaction events follow a Poisson distribution. In particular, different choices of $\Delta t$ have no effect on the reaction statistics.

\subsection{Gillespie algorithm: fixed ligands}

Compared to mobile ligands, fixed ligands can have a different number of neighboring receptors and therefore a different affinity to form a bridge. We define the affinity of ligand $i$ to form a bridge, $a_\mathrm{on}(i)$, and the total affinity of forming a bridge, $a_\mathrm{on}^T$, as
\begin{eqnarray} 
a_\mathrm{on}(i)=\kon n_R(i) \, , 
&\qquad &
a_\mathrm{on}^T=\sum_i a_\mathrm{on}(i) \, .
\end{eqnarray}
$a_\mathrm{off}^T$ is defined as in the case of mobile ligands (see previous section).
The chart flow of the program for firing reactions with fixed ligands is as follows 
\vspace{0.15 cm} \\
1 $t_\mathrm{reac}=0$\\
2 while$(t_\mathrm{reac}<\Delta t)$:\\
3 \hspace{0.5 cm} $t_\mathrm{reac}+=-\log(1-u_1)/a^T$   \\
4 \hspace{0.5 cm} if$(t_\mathrm{reac}<\Delta t)$:\\
5 \hspace{1. cm} if$(u_2<a_\mathrm{on}^T/a^T)$: \\
6 \hspace{1.5 cm} select a ligand $i$ with probability $p_i={a_\mathrm{on}(i)\over a_\mathrm{on}^T}$ \\
7 \hspace{1.5 cm} select a receptor $j$ between the $n_R(i)$ ones that could bind $i$ \\
8  \hspace{1.5 cm} bind $i$ with $j$ \\
9 \hspace{1.5 cm} update the affinity lists: $a_\mathrm{on}$, $a_\mathrm{off}$  \\
10 \hspace{1. cm} else: \\
11 \hspace{1.5 cm} unbind one of the bound receptors \\
12 \hspace{1.5 cm} update the affinity lists: $a_\mathrm{on}$, $a_\mathrm{off}$  \\
13 \# $u_i$: independent, uniform random numbers, $u_i\in[0,1)$

\section{Emerging diffusion constant at constant $\kon$}

Main Eqs.~6 and 10 report the emerging diffusion constant as function of $\koff$ and $n_b$. To predict the trends of data points obtained at a given $\kon$ and different $\koff$, we use the following relation (obtained in steady conditions):
\begin{eqnarray}
\kon n_r n^{(X)}_\ell = \koff \langle n_b\rangle &\qquad& X=\mathrm{fixed},\,\, \mathrm{mobile} \, ,
\label{Eq:SIkon}
\end{eqnarray}
where $n_r$ is the number of free receptors covered by the disk and $n^{(X)}_\ell$ is the average number of free ligands reachable by a given receptor. In particular, $n^{\mathrm{(mobile)}}_\ell$ is the total number of free ligands while $n^{\mathrm{(fixed)}}_\ell$ is the number of free ligands in the neighborhood of a given receptor. In Eq.~\ref{Eq:SIkon}, the fact that $n^{\mathrm{(mobile)}}_\ell \gg n^{\mathrm{(fixed)}}_\ell$ is compensated by the fact that, for mobile ligands, $\kon$ accounts also for the entropic cost of localizing a ligand in the proximity of a fixed receptor (see the $\pi R^2$ term in the definition of $\koff$, Main Eqs.~9).

Using Eq.~\ref{Eq:SIkon}, we derive the following expression for the emerging diffusion constant expressed as a function of $\kon$
\begin{eqnarray}
D &=&  { \kon n_r n^{(X)}_\ell A \over  \langle n_b \rangle^3  } 
\end{eqnarray}
where $A=\pi R^2$ and $A=\pi \lambda^2$, respectively, for mobile and fixed ligands. From the previous equation we predict that $D\sim \langle n_b \rangle^{-3}$ only if the number of free ligands and receptors is not sensibly depleted by the bridges (e.g., $n_r=N_R-\langle n_b\rangle$, $N_R \gg \langle n_b\rangle$).